\documentstyle[12pt]{article}
\renewcommand{\theequation}{\thesection.\arabic{equation}}
\newlength{\extraspace}
\newlength{\extraspaces}
\newcommand{\be}{\begin{equation}
\addtolength{\abovedisplayskip}{\extraspaces}
\addtolength{\belowdisplayskip}{\extraspaces}
\addtolength{\abovedisplayshortskip}{\extraspace}
\addtolength{\belowdisplayshortskip}{\extraspace}}
\newcommand{\ee}{\end{equation}}
 
\newcommand{\ba}{\begin{eqnarray}
\addtolength{\abovedisplayskip}{\extraspaces}
\addtolength{\belowdisplayskip}{\extraspaces}
\addtolength{\abovedisplayshortskip}{\extraspace}
\addtolength{\belowdisplayshortskip}{\extraspace}}
\newcommand{\ea}{\end{eqnarray}}
 
\newcommand{\bas}{\begin{eqnarray*}
\addtolength{\abovedisplayskip}{\extraspaces}
\addtolength{\belowdisplayskip}{\extraspaces}
\addtolength{\abovedisplayshortskip}{\extraspace}
\addtolength{\belowdisplayshortskip}{\extraspace}}
\newcommand{\eas}{\end{eqnarray*}}
 
\newcounter{subequation}[equation]
\makeatletter
\expandafter\let\expandafter\reset@font\csname reset@font\endcsname
\def\subeqnarray{\arraycolsep1pt
    \def\@eqnnum\stepcounter##1{\stepcounter{subequation}
        {\reset@font\rm(\theequation\alph{subequation})}}\eqnarray}

\makeatother
 
\newenvironment{theorem}[1]
{\vspace{3mm}\noindent {\bf #1 :} }{\vspace{2mm}}
\newcommand{\bt}[1]{\begin{theorem}{#1}}
\newcommand{\et}{\end{theorem}}
 
\newcommand{\newsection}[1]{
\vspace{12mm}
\pagebreak[3]
\addtocounter{section}{1}
\setcounter{equation}{0}
\setcounter{subsection}{0}
 
\begin{flushleft}
{\large\bf \thesection. #1}
\end{flushleft}
\nopagebreak
\medskip
\nopagebreak}
 
\newcommand{\newsubsection}[1]{
\vspace{1cm}
\pagebreak[3]
 
\addtocounter{subsection}{1}
\noindent{ \bf \thesubsection. #1}
\nopagebreak
\vspace{2mm}
\nopagebreak}
 
\newcommand{\NP}[1]{Nucl.\ Phys.\ {\bf #1}}

\newcommand{\PR}[1]{Phys.\ Rev.\ {\bf #1}}

\begin{document}
%
\begin{titlepage}
%
%
\begin{flushright}
NTZ 20/97\\
July 1997
\end{flushright}
\vspace{1cm}
 
\begin{center}
{\Large {\bf Complete control of gauge parameter dependence}} \\[4mm]
{\Large {\bf in the Abelian Higgs model}}
{\makebox[1cm]{  }       \\[1.5cm]
{\bf Rainer H\"au\ss ling and Stephan Kappel}\\ [3mm]
{\small\sl Institut f\"ur Theoretische Physik und} \\
{\small\sl Naturwissenschaftlich-Theoretisches 
           Zentrum, Universit\"at Leipzig} \\
{\small\sl Augustusplatz 10/11, D-04109 Leipzig, Germany} \\[0.5cm]
}
\vspace{1.5cm}
 
{\bf Abstract}
\end{center}
\begin{quote}
We examine the dependence on all gauge parameters in the example of the Abelian Higgs
model by applying a general algebraic method which roots in an extension of the usual Slavnov-Taylor identity. This method automatically yields all information about the gauge parameter dependence of Green functions and therefore especially allows to control the range of ``good'' normalization conditions. In this context we show that the physical on-shell normalization conditions are in complete agreement with the restrictions dictated by the enlarged Slavnov-Taylor identity and that the coupling can be fixed in an easily handleable way on the Ward identity of local gauge invariance. As an application of the general method we also study the Callan-Symanzik equation and the renormalization group equation of the Abelian Higgs model.
\end{quote}
\vfill
\renewcommand{\thefootnote}{\arabic{footnote}}
\setcounter{footnote}{0}
\end{titlepage}
%
\newsection{Introduction}

The need to fix the gauge when quantizing a gauge theory perturbatively introduces a set of arbitrary gauge parameters into the action. Therefore one unavoidably has the task to control
the dependence of the theory on these gauge parameters. Especially, it has to be proven that
{\it physical} quantities indeed are gauge parameter independent. For instance, the gauge parameter independence of the S-matrix, already suggested in \cite{lee}, was proven in \cite{BRS2} for gauge theories that do not contain any massless particles due to a complete spontaneous breakdown of symmetry. This proof, however, relies on a special set of on-shell normalization conditions and also makes use of a rather complicated technical tool, namely the Wilson operator
product expansion. On the other hand, looking at pure gauge theories with massless gauge bosons, where the S-matrix does not exist, the gauge parameter independence of the
$\beta$-functions has been shown. This, however, solely has been achieved by explicitly refering to an invariant renormalization scheme \cite{KSZ}.\\[0.2cm]
In the standard model of electroweak interactions the prerequisites needed for the proofs of the
examples mentioned above are not fulfilled due to the masslessness of the photon and parity violation in the fermion sector. Hence the state of the art concerning the control of gauge parameter dependence is quite unsatisfactory and the necessity for having at hand a general (i.e. model- and scheme-independent) and easily manageable tool arises. Such a tool is given by the algebraic method first proposed in \cite{PS} which also allows for the control of gauge parameter dependence of single Green functions. As a preparatory step for similar investigations in the standard model this general method has been applied to the Abelian Higgs model in \cite{HK}. But in \cite{HK} attention was restricted to the dependence of the theory on one gauge parameter only. Among other things results proven in \cite{niel} by explicitly using an invariant scheme and special properties of the model could be reproduced in a model- and scheme-independent way.
The present paper, now, enlarges the considerations of \cite{HK} to the {\it full} control of gauge parameter dependence (i.e. the control of the dependence of the theory on {\it all} gauge parameters) and hence completes the treatment of \cite{HK} in this sense. Again in view of the application of the algebraic method to the rather complicated standard model, this model containing
quite a lot of gauge parameters, it seems to be instructive and in fact necessary to {\it completely} work out this method, applied in its {\it full} extent, in the simpler case of the Abelian Higgs model as a preliminary. The necessary prerequisites for an analogous discussion of gauge parameter dependence in the standard model are in the meanwhile available due to \cite{K1}.\\[0.2cm]
The algebraic method essentially roots in a certain extension of the ordinary BRS transformations: All the gauge parameters of the model now are allowed to transform under BRS into Grassmann variables. It then follows that constructing the Green functions in accordance with this enlarged BRS invariance also automatically yields all information about the gauge parameter dependence of the original Green functions, some of which are also used in the normalization conditions. Because these normalization conditions have to be chosen in agreement with the gauge parameter dependence of the theory (in order not to ruin, for instance, the gauge parameter independence of the S-matrix) we hence have at hand a powerful tool for controlling the range of allowed normalization conditions. In this context it turns out that the conditions of \cite{BRS2} and \cite{KSZ} just build special sets of adequate normalization conditions (c.f. section 5, \cite{PS},\cite{KS2}).\\[0.2cm]
The structure of the paper will be as follows: In a first part (including sections 2--7) we enlarge the results obtained in \cite{HK} to the case when all gauge parameters undergo BRS transformations. This first part therefore parallels the discussion of \cite{HK} thereby putting emphasis on the modifications arising in the general, present, case. Whenever the treatment 
is {\it completely} analogous to the one in \cite{HK} we will skip calculational details and refer
to \cite{HK}, but nevertheless the present paper is fully self-contained as far as the line of argument is concerned:  We start with a short recapitulation of the Abelian Higgs model (section 2)
and the method of BRS transforming gauge parameters (section 3).  In section 4 we will look for the solution of the classical approximation. This solution also leads to restrictions for the gauge parameter dependence of some of the free parameters of the model. Section 5 deals with the extension of these restrictions to higher orders of the perturbative expansion and shows the compatibility of the extended restrictions with physical on-shell normalization conditions. In sections 6 and 7 we prove global and local Ward identities. Section 7 also contains the discussion of an alternative and more elegant possibility for fixing the coupling.\\
In order to illustrate how far one can get with algebraic considerations alone and also to complete
the algebraically abstract treatment we construct in a second part 
(sections 8, 9) parametric differential equations, namely the Callan-Symanzik equation and the
renormalization group equation of the Abelian Higgs model. In this context we also discuss the dependence of the theory on the ghost mass.\\
Section 10 summarizes the results.

\newsection{The Abelian Higgs model}

\newcommand{\f}{\underline{\varphi}}
\newcommand{\fe}{\varphi_1}
\newcommand{\fz}{\varphi_2}
We start with a short presentation of the Abelian Higgs model, thereby emphasizing some aspects which will become relevant in the following.
The model consists of a doublet of scalar fields
$\f = (\fe, \fz)$ and a gauge field $A_\mu$ with an interaction, that
breaks $U(1)$ gauge invariance spontaneously. In conventional normalization it can
be described by the classical action
\begin{equation}
  \Gamma_{inv} = \int\left\{ -\frac 14 F_{\mu \nu}F^{\mu \nu} +
  \frac 12 (D_\mu \f)(D^\mu \f) - \frac 18\frac{m_H^2}{m^2}e^2
  \left(\fe^2 + 2\frac me \fe + \fz^2\right)^2\right\}
\end{equation}
with:
\begin{equation}
  F_{\mu \nu}\equiv \partial_\mu A_\nu -\partial_\nu A_\mu \; \; , \;  \;
  D_\mu \fe \equiv \partial _\mu \fe + eA_\mu \fz \; \; , \; \;
  D_\mu \fz \equiv \partial _\mu \fz - eA_\mu \left(\fe +\frac me\right)
\end{equation}
$\Gamma _{inv}$ respects $U(1)$ symmetry and the discrete symmetry of
charge conjugation, i.e.\ it is invariant under the $U(1)$ transformations
\begin{equation}\label{gl1.3}
  \delta _\omega A_\mu = \partial _\mu \omega \; \; ,
\; \;
  \delta _\omega \fe = -e\omega \fz \; \; ,
\; \;
  \delta _\omega \fz = e\omega \left(\fe + \frac me\right)
\end{equation}
and under charge conjugation
\begin{equation}
  CA_\mu = -A_\mu ,
\; \;
  C\fe =  \fe,
\; \;
  C\fz =  - \fz .
\end{equation}
The shift $\frac me$ of the field $\fe$ produces the mass $m$ for the
vector field $A_\mu $ and $\fe$ is the physical Higgs field with mass $m_H$,
whereas $\fz$ takes the role of the would-be Goldstone boson eaten up
by $A_\mu $.\\
In order to quantize the model the gauge has to be fixed. To this end we introduce an additional field, namely the auxiliary field $B$, with $\delta _\omega B=0$ and add a gauge fixing term
\begin{equation}
\label{gl1.4}
  \Gamma _{g.f.} = \int\left\{ \frac 12 \xi B^2
  + B(\partial A +\xi_A m \fz)\right\}.
\end{equation}
$\xi $ and $\xi _A$ denote the gauge parameters, and the t'Hooft gauge fixing
term $\int B\xi _Am\fz$ is necessary in order to avoid a non-integrable
infrared singularity in the $\langle\fz \fz\rangle$ propagator.
Of course, this gauge fixing term violates both the local and global gauge invariance
non-trivially:
\begin{equation}
  \delta_\omega\Gamma_{g.f.} = \int\left\{\omega\Box B + e\omega B\xi_A m
  \left(\fe +\frac me \right) \right\}
\end{equation}
To retain a symmetry one has to further enlarge the model by introducing the 
Faddeev-Popov $(\phi \pi)$ fields $c,\bar{c}$ and to extend local gauge 
transformations to BRS transformations:
\begin{equation}
  \begin{array}{rcl@{\qquad}rcl}
  sA_\mu & = & \partial_\mu c, & sc & = & 0, \\
  s\fe & = & -ec\fz, & s\fz & = & ec\left(\fe + \frac me \right), \\
  s\bar{c} & = & B, & sB &= & 0
  \end{array}
\end{equation}
Adding the $\phi \pi$-action,
\begin{equation}
  \Gamma_{\phi\pi} = \int\left\{-\bar c\Box c 
  - e\bar{c}\xi_A m \left(\fe + \frac me \right)c\right\},
\end{equation}
the BRS variation of $\Gamma_{\phi\pi}$ exactly cancels the BRS variation
of $\Gamma_{g.f.}$, and $\Gamma_{inv} + \Gamma_{g.f.} +\Gamma_{\phi\pi}$ is BRS symmetric.
The BRS symmetry is a powerful technical tool which is essential for the proof of renormalizability
and unitarity of the S-matrix.  It also defines the model in question in an implicit way (see below).\\
Finally, we have to care about the non-linear BRS transformations $s \varphi_i$ which are not well-defined in higher orders of perturbation theory due to their non-linearity. In order to circumvent this difficulty we couple these BRS variations to external fields $Y_i$ with $s Y_i = 0$ and add an external field part:
\begin{equation}
  \Gamma_{e.f.} = \int\{Y_1(s\fe)+Y_2(s\fz)\}
\end{equation}
The complete BRS invariant classical action is now given by:
\begin{equation}
\label{glcl}
  \Gamma_{cl} = \Gamma_{inv} + \Gamma_{g.f.} + \Gamma_{\phi \pi} + \Gamma_{e.f.}
\end{equation}
A further complication arises when looking at rigid and local gauge invariance in terms of Ward identities: In \cite{KS} it has been shown that a {\it proper} formulation of rigid and local gauge symmetry (to all orders) is achieved by complementing the gauge fixing by an doublet
of external scalar fields $\hat{\underline{\varphi}} = (\hat{\varphi}_1, \hat{\varphi}_2 )$:
\begin{equation}
\label{ggf}
  \Gamma_{g.f.} = \int \left\{ \frac{1}{2} \xi B^2 +
  B \partial A - e 
  B\Bigl( (\hat \varphi_1 -  \xi_A \frac me )\varphi_2  -
  \hat \varphi_2 ( \varphi_1 -  \hat \xi_A \frac me ) \Bigr) \right\} 
\end{equation}  
The doublet $\hat{\underline{\varphi}}$ transforms under $U(1)$ according to:
\begin{equation}
\label{efg}
  \delta _{\omega} \hat \varphi_1 = - e \omega \hat \varphi_2 \; , \;
  \delta _{\omega} \hat \varphi_2 = e \omega (\hat \varphi_1 -\xi _A \frac me)
\end{equation}
Then $\Gamma_{g.f.}$ (\ref{ggf}) is invariant under the global $U(1)$ transformations (\ref{gl1.3}),
(\ref{efg}) if we choose $\hat{\xi}_A = -1$, and the original gauge fixing (\ref{gl1.4}) is recovered for $\hat{\varphi}_i = 0 $.\\
The external fields $\hat{\varphi}_i$ are transformed under BRS into another doublet of external fields $\underline{q} = (q_1, q_2 )$:
\begin{equation}
  s\hat \varphi_i  = q_i \; , \; s q_i = 0 \; , \; i = 1,2
\end{equation}
The BRS invariance of the theory is expressed by the Slavnov-Taylor (ST) identity
\begin{equation}
  {\cal S}(\Gamma)\equiv \int\left\{ \partial_\mu c\frac{\delta\Gamma}{\delta A_\mu} +
   B\frac{\delta\Gamma}{\delta\bar c} + 
   \frac{\delta\Gamma}{\delta\underline Y} \frac{\delta\Gamma}{\delta\underline\varphi} +
   \underline q \frac{\delta\Gamma}{\delta\underline{\hat{\varphi}}}\right\}=0 \; \; .
   \label{STI}
\end{equation}
At the classical level $\Gamma$ is just the classical action $\Gamma_{cl}$,
whereas at
the quantum level $\Gamma$ denotes the vertex functional 
$\Gamma = \Gamma_{cl} + O(\hbar)$.
It can be proven that (\ref{STI}) together with 
appropriate normalization conditions, invariance under charge conjugation 
and the gauge condition (\ref{ggf}) uniquely defines the model  to all orders
of perturbation theory. This is in contrast to the standard model where in addition to the ST identity 
\begin{table}
\begin{center}
\begin{tabular}{l|c|c|c|c|c|c|c|c|c|c}
fields & $A_{\mu}$ & $B$ & $\tilde{\varphi}_1$ & $\tilde{\varphi}_2$ &
 $c$ & $\bar{c}$ & $Y_1$ & $Y_2$ & $q_1$ & $q_2$ \\ \hline
dim    & 1 & 2 & 1 & 1 & 0 & 2 & 3 & 3 & 1 & 1 \\ \hline
charge conj. & - & - & + & - & - & - & + & - & + & - \\ \hline 
$Q_{\phi \pi}$ & 0 & 0 & 0 & 0 & +1 & -1 & -1 & -1 & +1 & +1 
\end{tabular}
\end{center}
{{\sl Table 1}: Quantum numbers of the fields ($\tilde{\varphi}_i =
\varphi_i, \hat{\varphi}_i$)}
\end{table}
also rigid invariance, a local Ward identity and some consistency relations are needed for a full
algebraic characterization of the model \cite{K1}.\\
In a first step one has to look for the most general, field polynomial (i.e. classical) solution
$\Gamma_{cl}^{gen}$ of the ST identity (\ref{STI}) and the gauge condition (\ref{ggf}) which is invariant under charge conjugation, and to prove that it coincides with $\Gamma_{cl}$ after the application of appropriate normalization conditions. This procedure also yields information about all the free parameters of the theory. The most general solution was calculated in \cite{KS} and is presented in appendix A. The free parameters in $\Gamma_{cl}^{gen}$ are the usual field and coupling renormalizations $z_1, z_2, z_A, z_m, z_{m_H}, z_e$,
\begin{eqnarray}
\label{gl2.15}
  & \varphi_i \longrightarrow \sqrt{z_i} (\varphi_i - x_i \hat{\varphi}_i ) \; \; , \; \;
     A_{\mu} \longrightarrow \sqrt{z_A} A_{\mu} & \\
  & m \longrightarrow \sqrt{z_m} m \; \; , \; \;
     m_H \longrightarrow \sqrt{z_{m_H}} m_H \; \; , \; \;
     e \longrightarrow z_e e \; \; \; \; , & \nonumber
\end{eqnarray}
as well as the gauge parameters $\xi, \xi_A$, the parameter $\mu$ (see appendix A), $\hat{\xi}_A$ (which is prescribed by the global Ward identity, see (\ref{gl6.3})) and the two parameters $x_1, x_2$, which appear in the combination
\begin{equation}
\label{comb}
  \bar{\varphi}_i = \varphi_i - x_i \hat{\varphi}_i
\end{equation}
that replaces $\varphi_i$ in $\Gamma_{inv}$.\\
These parameters have to be fixed by normalization conditions in each order of the perturbative expansion. In the following we will choose (for reasons which will become clear later on) physical on-shell normalization conditions:
\begin{eqnarray}
\label{massnorm}
Re \; \Gamma_{\varphi_1 \varphi_1} (p^2 = m_H^2 ) = 0 &\mbox{fixes} &  z_{m_H} 
 \nonumber \\
\Gamma^T (p^2 = m^2) = 0 & \mbox{fixes} &   z_m   \\
\Gamma_{c \bar{c}} (p^2 = m_{\hbox{ghost}}^2) = 0 \; \; , \; \;
m_{\hbox{ghost}}^2 = \xi_A m^2 \; \; & \mbox{fixes} &
  \xi_A \nonumber
\end{eqnarray}
\begin{eqnarray}
\label{wavenorm}
\partial_{p^2} \Gamma^T (p^2 = m^2) = 1 & \mbox{fixes} & z_A \nonumber \\
Re \; \partial_{p^2} \Gamma_{\varphi_1 \varphi_1} (p^2 = m_H^2) = 1
  & \mbox{fixes} & z_1  \\
\partial_{p^2} \Gamma_{\varphi_2 \varphi_2} (p^2 = \kappa^2) = 1
  & \mbox{fixes} & z_2 \; \nonumber 
\end{eqnarray}
\begin{eqnarray}
\label{xnorm}
\Gamma_{Y_1 q_1} (p^2 = \kappa^2) = x_1^{(0)} & \mbox{fixes} & x_1 \nonumber \\
\Gamma_{Y_2 q_2} (p^2 = \kappa^2) = x_2^{(0)} & \mbox{fixes} & x_2
\end{eqnarray}
\begin{equation}
\label{vacnorm}
\Gamma _{\varphi_1}
 = 0 \quad\hbox{fixes} \quad \mu
\end{equation}
In (\ref{massnorm}), (\ref{wavenorm}) the transversal part of the vector 2-point function is given by:
\begin{equation}
\label{transvec}
 \Gamma _{A^\mu A^\nu} (p,-p) \equiv \Gamma_{\mu \nu} (p,-p) 
= (\eta_{\mu \nu} -
  \frac{p_{\mu} p_{\nu}}{p^2}) \Gamma^T (p^2) + \frac{p_{\mu} p_{\nu}}{p^2}
  \Gamma^L(p^2)
\end{equation}
It remains to give a normalization condition for the coupling $e$. Preliminary (see sections 5 and 7), we fix the coupling on the 3-point vertex function $\Gamma_{A_{\mu} \varphi_1 \varphi_2}$ at a 
normalization momentum $p_{norm}$:
\begin{equation}
\label{coupnorm}
\partial_{p_1^\nu} \Gamma_{A_\mu \varphi_1 \varphi_2 } 
(-p_1-p_2,p_1,p_2) \Big|_{\{p_i\} = p_{norm} } =  - i e \eta^{\mu \nu}
  \quad \mbox{fixes} \quad z_e
\end{equation}
It is easily checked that these normalization conditions when applied to the tree approximation $\Gamma_{cl}^{gen}$ (see appendix A) exactly yield $\Gamma_{cl}$ (\ref{glcl}) (if we set $x_i = 0$).

\newsection{Algebraic control of gauge parameter dependence}

We now want to turn to the proper subject of the present paper, namely the control of gauge
parameter dependence. To this end we first observe that at the level of the classical action
$\Gamma_{cl}$ (\ref{glcl}) the dependence on the two gauge parameters $\xi$ and $\xi_A$ is given by two BRS variations,
\begin{eqnarray}
\label{gl3.1}
& \partial_{\xi} \Gamma_{cl} = \frac{1}{2} \int B^2 = \frac{1}{2} s \int \bar{c} B \; \; \;
   \mbox{ and} & \nonumber \\
& \partial_{\xi_A} \Gamma_{cl} = m \int \left\{ B \varphi_2 - e \bar{c} (\varphi_1 + \frac{m}{e}) c
   \right\} = m \; s \int \bar{c} \varphi_2 \; \; \; , &
\end{eqnarray}
respectively. Therefore the right hand sides of (\ref{gl3.1}) vanish between physical states and
physical quantities (like the S-matrix) are $\xi$- and $\xi_A$-independent in the tree
approximation.\\
The question now arises whether -- and if yes, how -- this statement can be extended to higher orders. In the affirmative case we furthermore would like 
to use a construction which is easily manageable and which does not rely on the specific
model and/or a specific renormalization scheme. Of course, if the model in question permits a
gauge-invariant regularization, such a general approach does not seem to be
necessary at first sight. But because
many models lack this property it is nevertheless desirable to have in hand such a model- and
scheme-independent procedure for controlling gauge parameter dependence and
to see how it works. In addition, it will turn out that some quite general results are {\it only} (or at
least much more easily) accessible with the proposed method.\\
For this purpose let us therefore allow the gauge parameters 
$\xi$ and $\xi_A$ to transform under BRS
into Grassmann variables $\chi$ and $\chi_A$, respectively, with $\phi \pi$-charge $+1$ \cite{PS}:
\begin{equation}
\label{gl3.2}
  s \xi = \chi \; \; \; , \; \; \; s \xi_A = \chi_A \; \; \; , \; \; \;
  s \chi = 0 = s \chi_A
\end{equation}
Hence, the ST identity modifies into:
\begin{equation}
\label{gl3.3}
  {\cal S} (\Gamma ) + \chi \partial_{\xi} \Gamma + \chi_A \partial_{\xi_A} \Gamma = 0
\end{equation}
Differentiation of (\ref{gl3.3}) with respect to $\chi$ or $\chi_A$ and evaluating the results at
$\chi = 0 = \chi_A$ leads to
\begin{eqnarray}
\label{gl3.4}
  - \left. s_{\Gamma}^{\chi = 0 = \chi_A} (\partial_{\chi} \Gamma ) \right|_{\chi = 0 = \chi_A} +
  \left. \partial_{\xi} \Gamma \right|_{\chi = 0 = \chi_A} & = & 0  \; \; \; , \nonumber \\
  - \left. s_{\Gamma}^{\chi = 0 = \chi_A} (\partial_{\chi_A} \Gamma ) \right|_{\chi = 0 = \chi_A} +
  \left. \partial_{\xi_A} \Gamma \right|_{\chi = 0 = \chi_A} & = & 0 \; \; \; ,
\end{eqnarray}
where in the model under investigation $s_{\Gamma}$ is given by:
\begin{equation}
\label{gl3.5}
  s_{\Gamma} = \int \left\{ \partial c \frac{\delta}{\delta A} +
                                         B \frac{\delta}{\delta \bar{c}} +
		\frac{\delta \Gamma}{\delta \underline{Y}} \frac{\delta}{\delta \underline{\varphi}} +
		\frac{\delta \Gamma}{\delta \underline{\varphi}} \frac{\delta}{\delta \underline{Y}} +
		\underline{q} \frac{\delta}{\delta \underline{\hat{\varphi}}} \right\} +
		\chi \partial_{\xi} + \chi_A \partial_{\xi_A}
\end{equation}
$s_{\Gamma}$ being -- roughly speaking -- the functional generalization of $s$, eq. (\ref{gl3.4}) is
nothing else but the functional analog of (\ref{gl3.1}) which we were looking for and which can be easily controlled in higher orders. Therefore proving (\ref{gl3.3}) to all orders of perturbation theory automatically yields all information about gauge parameter dependence of the 1-PI Green functions
in an algebraic way.

\newsection{Slavnov-Taylor identity for $\chi \neq 0$, $\chi_A \neq 0$}

In accordance with the observations of the proceeding section gauge parameter dependence
is completely governed by the $\chi$- and $\chi_A$-enlarged ST identity\footnote{From here on the symbol ${\cal S}$ collectively denotes all the differential operators on the r.h.s. of
(\ref{gl3.3})}:
\begin{equation}
\label{gl4.1}
  {\cal S}(\Gamma) \equiv \left\{ 
  \partial_{\mu} c \frac{\delta \Gamma}{\delta A_{\mu}} +
  B \frac{\delta \Gamma}{\delta \bar{c}} +
  \frac{\delta \Gamma}{\delta \underline{Y}}
    \frac{\delta \Gamma}{\delta \underline{\varphi}} +
  \underline{q} \frac{\delta \Gamma}{\delta \underline{\hat{\varphi}}}
  \right\}
  + \chi \partial_{\xi} \Gamma
  + \chi_A \partial_{\xi_A} \Gamma \; = \; 0
\end{equation}
First we have to look for the general {\it classical} solution $\Gamma = \Gamma_{cl}^{gen}$ of
(\ref{gl4.1}) in order to control the free parameters of the theory and to learn something about
their gauge parameter dependence eventually. Because the ST identity does not prescribe the gauge fixing terms we can also postulate the gauge condition (\ref{ggf})
\begin{equation}
\label{gl4.2}
  \left. \frac{\delta \Gamma}{\delta B} \right|_{\chi = 0 = \chi_A} =
  \xi B + \partial A -
  e\left[ (\hat{\varphi}_1 - \xi_A \frac{m}{e}) \varphi_2 -
               \hat{\varphi}_2 (\varphi_1 - \hat{\xi}_A \frac{m}{e}) \right]
\end{equation}
to hold for the solution $\Gamma$ of (\ref{gl4.1}). The gauge condition (\ref{gl4.2}) is linear in
propagating fields and hence it can be integrated in this form to all orders of perturbation theory.\\
Using the fact that $\chi$ and $\chi_A$ are Grassmann variables, $\Gamma$ can be split into four
parts in the tree approximation:
\begin{equation}
\label{gl4.3}
  \Gamma = \hat{\Gamma} + \chi Q + \chi_A Q_A + \chi \chi_A Q_{\chi \chi_A}
\end{equation}
Inserting (\ref{gl4.3}) into the ST identity (\ref{gl4.1}) and again making use of $\chi^2 = 0 =
\chi_A^2$ one immediately finds that at the classical level (\ref{gl4.1}) is equivalent to the following
four equations:
\begin{eqnarray}
\label{gl4.4}
  \chi^0, \chi_A^0 & \; \; : \; \;&  \int \left\{ \partial_{\mu} c 
     \frac{\delta \hat{\Gamma}}{\delta A_{\mu}} +
     B \frac{\delta \hat{\Gamma}}{\delta \bar{c}} +
     \frac{\delta \hat{\Gamma}}{\delta \underline{Y}}
     \frac{\delta \hat{\Gamma}}{\delta \underline{\varphi}} +
     \underline{q} \frac{\delta \hat{\Gamma}}{\delta \underline{\hat{\varphi}}}
     \right\} \; \;  = \; \; 0 \\
 \label{gl4.5}
  \chi^1, \chi_A^0 & \; \; : \; \;& \partial_{\xi} \hat{\Gamma} \; \;  = \; \; 
                                     s_{\hat{\Gamma}}^{\chi = 0 =\chi_A} Q \\
\label{gl4.6}
  \chi^0, \chi_A^1 & \; \; : \; \;& \partial_{\xi_A} \hat{\Gamma} \; \;  = \; \;  
                                     s_{\hat{\Gamma}}^{\chi = 0 = \chi_A} Q_A \\
\label{gl4.7}
  \chi^1, \chi_A^1 & \; \; : \; \;& \int \left\{ \frac{\delta Q}{\delta \underline{Y}}
                  \frac{\delta Q_A}{\delta \underline{\varphi}} -
      \frac{\delta Q_A}{\delta \underline{Y}}
      \frac{\delta Q}{\delta \underline{\varphi}} \right\} -
      \partial_{\xi} Q_A + \partial_{\xi_A} Q \; \; = \; \;
      s_{\hat{\Gamma}}^{\chi = 0 = \chi_A} Q_{\chi \chi_A}
\end{eqnarray}
$s_{\hat{\Gamma}}$ is given by (\ref{gl3.5}) (with $\hat{\Gamma}$ replacing $\Gamma$).\\
The first of these equations is nothing else but the (ordinary) ST identity for $\chi = 0 = \chi_A$ which has been studied in \cite{KS} and the general solution 
of which -- needed for the calculation of
$Q$ and $Q_A$ -- is presented in appendix A.\\
Furthermore, (\ref{gl4.3}) implies that $Q$, $Q_A$ and $Q_{\chi \chi_A}$ have dimension less than
or equal to four and are even under charge conjugation and that $Q$ as well as $Q_A$
carry $\phi \pi$-charge $-1$ whereas $Q_{\chi \chi_A}$ has $\phi \pi$-charge $-2$. According
to the table of quantum numbers the most general ansatz for $Q$ is hence given by:
\begin{eqnarray}
\label{gl4.8}
  Q & = &  \int \left\{ d_1 Y_1 \varphi_1 + \hat{d}_1 Y_1 \hat{\varphi}_1 + d Y_1 +
                                d_2 Y_2 \varphi_2 + \hat{d}_2 Y_2 \hat{\varphi}_2 \right. \nonumber \\
      &    & \; \; \; \left. f \bar{c} \varphi_2 + \hat{f} \bar{c} \hat{\varphi}_2 +
                         \tilde{f} \bar{c} B + f_A \bar{c} \partial_{\mu} A^{\mu} \right. \nonumber \\
      &    & \; \; \; \left. h_1 \bar{c} \varphi_1 \varphi_2 + h_2 \bar{c} \hat{\varphi}_1 \varphi_2 +
                         h_3 \bar{c} \varphi_1 \hat{\varphi}_2 +
			  h_4 \bar{c} \hat{\varphi}_1 \hat{\varphi}_2 \right\}
\end{eqnarray}
For $Q_A$ the same ansatz holds true but with a new set of $13$ parameters 
$d_1^A, \hat{d}_1^A, \dots , h_4^A$ instead of $d_1, \hat{d}_1, \dots , h_4$. 
Due to the quantum numbers of $Q_{\chi \chi_A}$ there
are no terms contributing to $Q_{\chi \chi_A}$:
\begin{equation}
\label{gl4.9}
  Q_{\chi \chi_A} \equiv 0
\end{equation}
Putting (\ref{gl4.8}) and the analogous expression for $Q_A$ into (\ref{gl4.5}), (\ref{gl4.6}),
respectively, yields after a straightforward calculation the determination of the 26 parameters
$d_1, \dots , h_4, d_1^A, \dots$, $h_4^A$; we finally find ($Q_{(A)} = Q, Q_A$):
\begin{equation}
\label{gl4.10}
  Q_{(A)} = Q_{e.f.(A)} + Q_{\phi \pi ,1(A)} + Q_{\phi \pi ,2(A)}
\end{equation}
with ($x_1^{(0)}  = x_2^{(0)} \equiv x$ (see (\ref{gl6.3})) and $\bar{\varphi}_i = \varphi_i - x \hat{\varphi}_i$):
\begin{eqnarray}
\label{gl4.11}
  Q_{e.f.(A)} & = & \int \left\{ \frac{1}{4} (\partial_{\xi_{(A)}} \mbox{ln} z_1 +
                                                           \partial_{\xi_{(A)}} \mbox{ln} z_2 )
							   (Y_1 \bar{\varphi}_1 + Y_2 \bar{\varphi}_2 ) \right. \nonumber \\
  & & \! \! \! \! \left. + \frac{1}{4} (\partial_{\xi_{(A)}} \mbox{ln} z_1 -
                                                  \partial_{\xi_{(A)}} \mbox{ln} z_2 )
                                          (Y_1 \bar{\varphi}_1 - Y_2 \bar{\varphi}_2 ) -
	  \partial_{\xi_{(A)}} x (Y_1 \hat{\varphi}_1 + Y_2 \hat{\varphi}_2 ) \right\} \\
  Q_{\phi \pi ,1(A)} & = & \int \left\{ -\frac{1}{4} e \bar{c}
                                         (\partial_{\xi_{(A)}} \mbox{ln} z_1 +
                                          \partial_{\xi_{(A)}} \mbox{ln} z_2 )
  \left( (\bar{\varphi}_1 + \frac{\sqrt{z_m}}{\sqrt{z_1} z_e} \frac{m}{e} ) \hat{\varphi}_2 -
          \bar{\varphi}_2 (\hat{\varphi}_1 - \xi_A \frac{m}{e} ) \right) \right. \nonumber \\
  & & \! \! \! \! \left. -\frac{1}{4} e \bar{c}
                     (\partial_{\xi_{(A)}} \mbox{ln} z_1 -
                      \partial_{\xi_{(A)}} \mbox{ln} z_2 )
  \left( (\bar{\varphi}_1 + \frac{\sqrt{z_m}}{\sqrt{z_1} z_e} \frac{m}{e} ) \hat{\varphi}_2 +
          \bar{\varphi}_2 (\hat{\varphi}_1 - \xi_A \frac{m}{e} ) \right) \right\} \\
  Q_{\phi \pi ,2} & = & \frac{1}{2} \int \bar{c} B  \\
  Q_{\phi \pi ,2 \; A} & = & m \int \bar{c} \bar{\varphi}_2
\end{eqnarray}
Please note that with (\ref{gl4.10}) eq. (\ref{gl4.7}) is fulfilled automatically.\\
Hence the coefficients in $Q$ and $Q_A$ are completely determined as functions of the
parameters $z_1, z_2, z_m, z_e$ and $x$ which appear in the general solution of the
ST identity for $\chi = 0 = \chi_A$. But the $\chi$- and $\chi_A$-enlarged ST identity does not only fully fix $Q$ and $Q_A$; in addition (\ref{gl4.5}), (\ref{gl4.6}) force some of the free
parameters to be both $\xi$- and $\xi_A$-independent:
\begin{eqnarray}
\label{gl4.15}
  & \partial_{\xi} z_e = 0 = \partial_{\xi_A} z_e \; \; , \; \;
      \partial_{\xi} z_A = 0 = \partial_{\xi_A} z_A \; \; , & \\
   & \partial_{\xi} z_m = 0 = \partial_{\xi_A} z_m \; \; , \; \;
       \partial_{\xi}z_{m_H} = 0 = \partial_{\xi_A} z_{m_H} \; \; , \; \;
       \partial_{\xi} \mu^2 = 0 = \partial_{\xi_A} \mu^2 & \nonumber
\end{eqnarray}
In contrast to this the wave function renormalizations $z_1, z_2$ and $x$ can be arbitrary functions of
$\xi$ and $\xi_A$.\\
Two remarks are of some relevance at this point:\\
The (physical) normalization conditions given in section 2 trivially fulfil the constraints (\ref{gl4.15}) in the tree approximation. 
In higher orders of perturbation theory, however, the constraints (\ref{gl4.15}) will extend to restrictions of the $\xi$- and 
$\xi_A$-dependence of some non-local Green functions (the subject of the next section) which
are also used in the normalization conditions and the splitting of which into $\xi (\xi_A)$-dependent and $\xi (\xi_A)$-independent parts is much less transparent. Hence some care is needed in order not to introduce wrong gauge parameter dependence into the theory, i.e. it has to be proven
explicitly that the normalization conditions chosen are in agreement with the restrictions
(\ref{gl4.15}) extended to higher orders.\\
The second remark concerns the t`Hooft gauge
\begin{equation}
\label{gl4.16}
  \xi_A = \xi
\end{equation}
which seems to be excluded in the present treatment because $\xi$ and $\xi_A$ are viewed
as being {\it independent} gauge parameters. But with the following recipe it is nevertheless
possible to make a transition from the general to the t`Hooft case:
\begin{itemize}
\item
  Set $\partial_{\xi_A}$ equal to zero in all places of occurence, this partial derivative having                  
  already been taken into account in the t`Hooft gauge via $s \xi = \chi$
\item
  Take then $\chi_A = \chi$
\end{itemize}
It is easily seen that this procedure leads to the correct results.

\newsection{Gauge parameter dependence of Green functions}

The next step would be the proof of the $\chi$- and $\chi_A$-dependent ST identity
(\ref{gl4.1}) to all orders of perturbation theory. We will not present the detailed proof here
but instead refer to \cite{PS} where it was shown that the proof of the enlarged ST identity
($\chi \neq 0, \chi_A \neq 0$) can be reduced to the proof of the ordinary ST identity
($\chi = 0 = \chi_A$): The only possible obstruction to the validity of the ST identity would be
the presence of anomalies which, however, are absent in the Abelian Higgs model. Hence we
can acchieve
\begin{equation}
\label{gl5.1}
  {\cal S} (\Gamma ) = 0
\end{equation}
also in the case of BRS transforming gauge parameters $\xi$ and $\xi_A$, namely by an
appropriate choice of counterterms. $\Gamma$ now denotes the generating functional of
1-PI Green functions. Accordingly the validity of (\ref{gl5.1}) will be assumed throughout the
following.\\[0.2cm]
We now want to deal with the extensions of the constraints (\ref{gl4.15}) to higher 
orders\footnote{In this context we will restrict ourselfes to the case of a stable Higgs particle, i.e.
$m_H^2 < 4 m^2$.}. Because this discussion again parallels the analogous discussion of \cite{HK} for one
BRS transforming gauge parameter we skip the details of the calculations here.\\
The fundamental starting point for all considerations that follow are the equations (\ref{gl3.4}) which have to be differentiated with respect to suitable fields and finally evaluated for all fields equal
to zero.\\[0.2cm]
Let us start with a more technical point: the continuation of the $\xi$- and $\xi_A$-independence
of $\mu^2$ to higher orders. To this end we differentiate (\ref{gl3.4}) with respect to $\varphi_1$; in momentum space we get:
\begin{equation}
\label{gl5.2}
  \partial_{\chi_{(A)}} \Gamma_{Y_1} (0) \; \; \Gamma_{\varphi_1 \varphi_1} (0) +
  \partial_{\chi_{(A)}} \Gamma_{Y_1 \varphi_1} (0) \; \; \Gamma_{\varphi_1} (0) =
  - \partial_{\xi_{(A)}} \Gamma_{\varphi_1} (0)
\end{equation}
Using the normalization condition $\Gamma_{\varphi_1} = 0$ (\ref{vacnorm}), eq. (\ref{gl5.2}) simplifies to
\begin{equation}
\label{gl5.3}
  \partial_{\chi_{(A)}} \Gamma_{Y_1} (0) \left( -m_H^2 + {\cal O} (\hbar) \right) = 0 \; \; \; ,
\end{equation}
from which it follows that
\begin{equation}
\label{gl5.4}
  \partial_{\chi} \Gamma_{Y_1} = 0 \; \; \mbox{ and } \; \;
  \partial_{\chi_A} \Gamma_{Y_1} = 0
\end{equation}
hold to all orders of the perturbative expansion.\\[0.2cm]
We next come to the proof of the statement that the transversal part of the vector 2-point function is completely gauge parameter independent (Classically this statement is true due to
$\partial_{\xi} z_A = 0 = \partial_{\xi_A} z_A$ and $\partial_{\xi} z_m = 0 =
\partial_{\xi_A} z_m$.): Differentiation of (\ref{gl3.4}) with respect to $A_{\mu}$ and $A_{\nu}$ leads
to (thereby using (\ref{gl5.4})):
\begin{equation}
\label{gl5.5}
  \partial_{\chi_{(A)}} \Gamma_{Y_2 A_{\mu}} (p,-p) \; \; \Gamma_{\varphi_2 A_{\nu}} (p,-p) +
  (\mu \leftrightarrow \nu ) =
  - \partial_{\xi_{(A)}} \Gamma_{A_{\mu} A_{\nu}} (p,-p)
\end{equation}
A simple argument using Lorentz invariance (see also \cite{HK}) shows that the left hand sides of (\ref{gl5.5}) only contribute to the longitudinal part of $\partial_{\xi} \Gamma_{A_{\mu} A_{\nu}}$ and
$\partial_{\xi_A} \Gamma_{A_{\mu} A_{\nu}}$, respectively. Therefore we get the desired
result:
\begin{equation}
\label{gl5.6}
  \partial_{\xi} \Gamma_{A_{\mu} A_{\nu}}^T = 0 \; \; \mbox{ and } \; \;
  \partial_{\xi_A} \Gamma_{A_{\mu} A_{\nu}}^T = 0
\end{equation}
(The transversal part of $\Gamma_{A_{\mu} A_{\nu}}$ is defined in (\ref{transvec}).)
Finally it is easy (but nevertheless {\it necessary}) to prove that the on-shell normalization
conditions (\ref{massnorm}), (\ref{wavenorm}) involving $\Gamma_{A_{\mu} A_{\nu}}^T$ 
are in agreement with the constraints (\ref{gl5.6}).\\
Please also note that the restrictions found above for the transversal part of the vector
2-point function are only available in this simple way by controlling gauge parameter
dependence algebraically.\\[0.2cm]
In a quite analogous manner the constraint $\partial_{\xi} z_{m_H} = 0 = \partial_{\xi_A} z_{m_H}$ is extended to higher orders: This time we differentiate (\ref{gl3.4}) twice with respect to
$\varphi_1$ (and use again (\ref{gl5.4})):
\begin{equation}
\label{gl5.7}
  \partial_{\chi_{(A)}} \Gamma_{Y_1 \varphi_1} (p^2) \; \; 
  \Gamma_{\varphi_1 \varphi_1}  (p^2) = - \partial_{\xi_{(A)}} 
  \Gamma_{\varphi_1 \varphi_1} (p^2)
\end{equation}
Equation (\ref{gl5.7}) completely governs the $\xi$- and $\xi_A$-dependence of the Higgs self-energy. But due to the existence of non-trivial insertions of the 
vertices $\chi \bar{c} B$ and $\chi_A m \bar{c} \varphi_2$ into the vertex function $\Gamma_{Y_1 \varphi_1}$ the l.h.s of (\ref{gl5.7}) is not trivial at all, this being in contrast to the discussion of the transversal part of the vector 2-point function. Nevertheless, it is easily shown (order by order in perturbation theory) that the on-shell normalization condition (\ref{massnorm}) is in agreement with the constraint (\ref{gl5.7}). (See \cite{HK} for a more detailed discussion.)\\[0.2cm]
We want to conclude this section by making some remarks concerning the extension of the
constraints $\partial_{\xi} z_e = 0 = \partial_{\xi_A} z_e$ to higher orders. Testing (\ref{gl3.4})
with respect to $A_{\mu}, \varphi_1$ and $\varphi_2$ leads to:
\begin{eqnarray}
\label{gl5.10}
  \partial_{\chi_{(A)}} \Gamma_{Y_1 \varphi_1} (p_1^2) \; \;
  \Gamma_{\varphi_1 \varphi_2 A_{\mu}} (p_1, p_2, p) & + &
  \partial_{\chi_{(A)}} \Gamma_{Y_2 \varphi_2} (p_2^2) \; \;
  \Gamma_{\varphi_1 \varphi_2 A_{\mu}} (p_1, p_2, p) \nonumber \\
  + \; \; \partial_{\chi_{(A)}} \Gamma_{Y_2 \varphi_1 A_{\mu}} (p_2, p_1, p) \; \;
  \Gamma_{\varphi_2 \varphi_2} (p_2^2) & + &
  \partial_{\chi_{(A)}} \Gamma_{Y_1 \varphi_2 A_{\mu}} (p_1, p_2, p) \; \;
  \Gamma_{\varphi_1 \varphi_1} (p_1^2) \\
  + \; \; \partial_{\chi_{(A)}} \Gamma_{Y_2 \varphi_1 \varphi_2} (p, p_1, p_2) \; \;
  \Gamma_{\varphi_2 A_{\mu}} (-p, p) & = &
  - \; \; \partial_{\xi_{(A)}} \Gamma_{\varphi_1 \varphi_2 A_{\mu}} (p_1, p_2, p) \nonumber
\end{eqnarray}
The left hand sides of (\ref{gl5.10}) do not contain any free parameters once the residua of the
Higgs and the would-be Goldstone are fixed by the normalization conditions (\ref{wavenorm}).
Hence (\ref{gl5.10}) completely determines the $\xi$- and $\xi_A$-dependence of
the vertex $\Gamma_{A_{\mu} \varphi_1 \varphi_2}$. If one therefore insists in fixing the coupling directly with the help of $\Gamma_{A_{\mu} \varphi_1 \varphi_2}$
(as it was done in (\ref{coupnorm})) one has to introduce
two reference points $\xi_0$ and $\xi_{A_0}$ in order to fix the $\xi$- and $\xi_A$-independent
part of $\Gamma_{A_{\mu} \varphi_1 \varphi_2}$,
\begin{equation}
\label{gl5.11}
  \left. \partial_{p_1^{\nu}} \Gamma_{A_{\mu} \varphi_1 \varphi_2} (- p_1 - p_2, p_1, p_2)
  \right|_{\{p_i\} = p_{norm}, \xi_ = \xi_0, \xi_A = \xi_{A_0}} =
  - i e \eta^{\mu \nu} \; \; \; ,
\end{equation}
and to govern $\xi$- and $\xi_A$-dependence via (\ref{gl5.10}). Such a procedure, however, is
not evident and easily manageable in explicit calculations at all. In section 7 we will see that in the Abelian Higgs model there is a much more elegant and practicable way of fixing the coupling, namely by making use of the local Ward identity.

\newsection{Rigid invariance}

In \cite{KS} it was proven that the $\chi$- and $\chi_A$-independent part of the generating functional of 1-PI Green functions obeys a Ward identity of rigid symmetry to all orders of perturbation theory,
\begin{equation}
\label{gl6.1}
  \hat{W}^{gen} \left. \Gamma \right|_{\chi = 0 = \chi_A} = 0 \; \; \; ,
\end{equation}
where $\hat{W}^{gen}$ denotes the (deformed) Ward operator:
\begin{eqnarray}
\label{gl6.2}
  \hat{W}^{gen} & \equiv & \int \left\{
  - z^{-1} \varphi_2 \frac{\delta}{\delta \varphi_1} +
  z (\varphi_1 - \hat{\xi}_A \frac{m}{e} ) \frac{\delta}{\delta \varphi_2} -
  z Y_2 \frac{\delta}{\delta Y_1} +
  z^{-1} Y_1 \frac{\delta}{\delta Y_2} \right. \nonumber \\
  & & \; \; \; \; \left. - z^{-1} \hat{\varphi}_2 \frac{\delta}{\delta \hat{\varphi}_1} +
  z (\hat{\varphi}_1 - \xi_A \frac{m}{e} ) \frac{\delta}{\delta \hat{\varphi}_2} -
  z^{-1} q_2 \frac{\delta}{\delta q_1} +
  z q_1 \frac{\delta}{\delta q_2} \right\}
\end{eqnarray}
The appearance of a {\it deformed} Ward operator is due to the fact that physical on-shell normalization conditions (which are ``good'' normalization conditions, see section 5) have been used. In other words: The WI (\ref{gl6.1}) does not prescribe the values of $z$ and
$\xi_A$, instead these parameters are fixed uniquely by explicit normalization conditions, namely the normalization conditions imposed on the residua of the Higgs and Goldstone field (\ref{wavenorm}) and the mass normalization of the ghosts and the Higgs (\ref{massnorm}).\\
Nevertheless, (\ref{gl6.1}) restricts some other parameters {\it at the classical level}:
\begin{equation}
\label{gl6.3}
  x_1^{(0)} = x_2^{(0)} \equiv x \; \; , \; \; \hat{\xi}_A = -1 + x \xi_A
\end{equation}
Now we are going to study the modifications of (\ref{gl6.1}) when BRS transforming gauge parameters $\xi$ and $\xi_A$ are included. We will start with a more detailed investigation
of the classical approximation, these considerations yielding a hint of what could be expected in higher orders, then we will outline the essential steps for the proof of the WI obtained classically
to all orders of perturbation theory.\\[0.3cm]
Acting with $\hat{W}^{gen}$ (\ref{gl6.2}) on the general solution $\Gamma_{cl}^{gen}$ (\ref{gl4.3}) of the ST identity (\ref{gl4.1}) yields (using $\hat{W}^{gen} \hat{\Gamma}_{cl}^{gen} = 0$):
\begin{equation}
\label{gl6.4}
  \hat{W}^{gen} \Gamma_{cl}^{gen} =
  \hat{W}^{gen} (\hat{\Gamma}_{cl}^{gen} + \chi Q + \chi_A Q_A ) =
  \chi \hat{W}^{gen} Q + \chi_A \hat{W}^{gen} Q_A
\end{equation}
The r.h.s. of (\ref{gl6.4}) is not only non-vanishing, but even worse it contains terms which are
non-linear in the propagating fields (terms proportional to $\bar{c} \varphi_1, \bar{c} \varphi_2$):
\begin{eqnarray}
\label{gl6.5}
  \hat{W}^{gen} Q_{(A)} & = &
  \int \left\{ - \partial_{\xi_{(A)}} z \; \left( Y_2 + e \bar{c} (\hat{\varphi}_1 - \xi_A \frac{m}{e} ) \right)
  \left( \bar{\varphi}_1 + \frac{\sqrt{z_m}}{\sqrt{z_1} z_e} \frac{m}{e} \right)
  \right. \nonumber \\
  & & \; \; \; \; \left. +\partial_{\xi_{(A)}} z^{-1} \; (Y_1 - e \bar{c} \hat{\varphi}_2 ) 
  \bar{\varphi}_2 \right. \\
  & & \; \; \; \; \left. + z \frac{m}{e} \left( (\partial_{\xi_{(A)}} \hat{\xi}_A - 
                                                             x \partial_{\xi_{(A)}} \xi_A ) Y_2 +
  (\partial_{\xi_{(A)}} \xi_A ) \; e \bar{c}
  \left( \bar{\varphi}_1 + \frac{\sqrt{z_m}}{\sqrt{z_1} z_e} \frac{m}{e} \right) \right) \right\} \nonumber
\end{eqnarray}
These non-linear terms are potentially harmful because they are not well-defined in higher orders.
In order to overcome this difficulty we will absorb the harmful terms -- in direct analogy to the
treatment of \cite{HK} -- into functional operators $\chi V^{gen}$ and $\chi_A V_A^{gen}$ which cancel these terms when acting on $\Gamma_{cl}^{gen}$. A natural choice for $V^{gen}$ and
$V_A^{gen}$ is given by\footnote{The expressions for $V^{gen}$ and $V_A^{gen}$ are most easily found by extending the operators in $\hat{W}^{gen}$ as far as possible to invariant operators of the $\chi$- and $\chi_A$-enlarged BRS transformations.}:
\begin{equation}
\label{gl6.6}
  V_{(A)}^{gen} = \partial_{\xi_{(A)}} \int \left\{
  z (\hat{\varphi}_1 - \xi_A \frac{m}{e} ) \frac{\delta}{\delta q_2} -
  z^{-1} \hat{\varphi}_2 \frac{\delta}{\delta q_1} \right\}
\end{equation}
An easy calculation now proves that the $\chi$- and $\chi_A$-enlarged Ward operator
\begin{equation}
\label{gl6.7}
  W^{gen} = \hat{W}^{gen} + \chi V^{gen} + \chi_A V_A^{gen} \; \; \; ,
\end{equation}
when acting on $\Gamma_{cl}^{gen}$, only leads to terms linear in the propagating fields:
\begin{equation}
\label{gl6.8}
  W^{gen} \Gamma_{cl}^{gen} = \chi \Delta_{br} + \chi_A \Delta_{br_A}
\end{equation}
with
\begin{equation}
\label{gl6.9}
  \Delta_{br_{(A)}} = \partial_{\xi_{(A)}} \int \left\{
  z^{-1} Y_1 \varphi_2 -
  z Y_2 (\varphi_1 - \hat{\xi}_A \frac{m}{e} ) \right\}
\end{equation}
The terms on the r.h.s. of (\ref{gl6.8}) are harmless because they cannot be inserted non-trivially 
into higher orders' loop diagrams. This concludes the classical treatment.\\[0.3cm]
Next we want to show that the WI (\ref{gl6.8}) is valid in all orders of perturbation theory:
\begin{equation}
\label{gl6.10}
  W^{gen} \Gamma = \chi \Delta_{br} + \chi_A \Delta_{br_A}
\end{equation}
$\Gamma$ now denotes the generating functional of 1-PI Green functions.\\  
Because this proof almost completely parallels the proof given in \cite{HK} for one BRS transforming gauge parameter, we will concentrate on the essential steps only and skip some calculational details in between.\\
In order to work scheme-independently as far as possible when proving (\ref{gl6.10}) we will only rely on the action principle whose validity has been shown in every renormalization scheme in use. This action principle implies:
\begin{equation}
\label{gl6.11}
  W^{gen} \Gamma = \tilde{\Delta} \cdot \Gamma
\end{equation}
$\tilde{\Delta}$ is a local (i.e. field polynomial) integrated insertion carrying the quantum numbers:
dim $\tilde{\Delta} \leq 4$, $C (\tilde{\Delta} ) : -$, $\phi \pi (\tilde{\Delta} ) = 0$.\\
The second ingredient, needed for the proof, is the transformation behaviour of $W^{gen}$
(\ref{gl6.7}) under BRS transformations,
\begin{eqnarray}
\label{gl6.12}
  0 & = & W^{gen} {\cal S} (\Gamma ) = s_{\Gamma} (W^{gen} \Gamma ) \\
  & & \! \! \! \! \! \! - \left[ (\chi \partial_{\xi} + \chi_A \partial_{\xi_A} ) \int \left\{
  -z^{-1} \varphi_2 \frac{\delta}{\delta \varphi_1} +
  z (\varphi_1 - \hat{\xi}_A \frac{m}{e} ) \frac{\delta}{\delta \varphi_2} -
  z Y_2 \frac{\delta}{\delta Y_1} +
  z^{-1} Y_1 \frac{\delta}{\delta Y_2} \right\} \right] \Gamma \; \; \; , \nonumber
\end{eqnarray}
where $s_{\Gamma}$ is given by (\ref{gl3.5}).\\
Furthermore, another straightforward calculation yields:
\begin{eqnarray}
\label{gl6.13}
  & s_{\Gamma} ( \chi \Delta_{br} + \chi_A \Delta_{br_A} ) = & \\ 
  & + \left[ ( \chi \partial_{\xi} + \chi_A \partial_{\xi_A} ) \displaystyle\int \left\{
  -z^{-1} \varphi_2 \displaystyle\frac{\delta}{\delta \varphi_1} +
  z (\varphi_1 - \hat{\xi}_A \displaystyle\frac{m}{e} ) \displaystyle\frac{\delta}{\delta \varphi_2} -
  z Y_2 \displaystyle\frac{\delta}{\delta Y_1} +
  z^{-1} Y_1 \displaystyle\frac{\delta}{\delta Y_2} \right\} \right] \Gamma & \nonumber
\end{eqnarray}
Hence combining (\ref{gl6.12}) and (\ref{gl6.13}) we find that the breaking of the WI (\ref{gl6.11}) has to be $s_{\Gamma}$-invariant:
\begin{equation}
\label{gl6.14}
  s_{\Gamma} (\tilde{\Delta} \cdot \Gamma - \chi \Delta_{br} - \chi_A \Delta_{br_A} ) = 0
\end{equation}
From here on the proof of the $\chi$- and $\chi_A$-enlarged Ward identity proceeds by induction in the loop expansion. In the tree approximation this WI has already been established (see above), and according to the action principle we have at 1-loop order:
\begin{equation}
\label{gl6.15}
  (W^{gen} \Gamma )^{(\leq 1)} = \tilde{\Delta}^{(1)}
\end{equation}
Using the validity of the WI for $\chi = 0 = \chi_A$ (\ref{gl6.1}) \cite{KS}, 
$\tilde{\Delta}^{(\leq 1)}$ must have the form:
\begin{equation}
\label{gl6.16}
  \tilde{\Delta}^{(1)} = \chi (\Delta_{br}^{(\leq 1)} + \Delta_-^{(1)}) + 
                                  \chi_A (\Delta_{br_A}^{(\leq 1)} + \Delta_{A -}^{(1)} )
\end{equation}
$\Delta_-^{(1)}, \Delta_{A -}^{(1)}$ are local insertions which carry $\phi \pi$-charge $-1$, and due to the quantum numbers of the fields in question no term proportional to $\chi \chi_A$ can appear in (\ref{gl6.16}).\\
The application of $s_{\Gamma}$ to (\ref{gl6.15}) then leads to
\begin{equation}
\label{gl6.17}
  s_{\Gamma} (\chi \Delta_-^{(1)} + \chi_A \Delta_{A -}^{(1)} ) =
  s_{\Gamma_{cl}} (\chi \Delta_-^{(1)} + \chi_A \Delta_{A -}^{(1)} ) + {\cal O} (\hbar^2 ) = 0
\end{equation}
and hence we rest with a purely classical cohomology problem. This classical problem is solved as usual: First one has to find two bases of field polynomials constituting $\Delta_-^{(1)}$ and
$\Delta_{A -}^{(1)}$, respectively ($v = \frac{m}{e}$):
\begin{eqnarray}
\label{gl6.18}
{\Delta}_{(A) -}^{(1)} 
& = & {\textstyle \int} \left\{ w_{(A) 1} Y_1 \varphi_2 +
  w_{(A) 2} Y_1 \hat{\varphi}_2 + w_{(A) 3} Y_2 + w_{(A) 4} Y_2 \varphi_1 +
  w_{(A) 5} Y_2 \hat{\varphi}_1 \right. \nonumber \\
& & \left. + w_{(A) 6} \bar{c} + w_{(A) 7} (-x Y_2 + e\bar{c} (\bar{\varphi}_1 + v)) 
    + w_{(A) 8} \bar{c} \hat{\varphi}_1
    + w_{(A) 9} \bar{c} A^2 \right. \nonumber \\
& & \left. + w_{(A) 10} \bar{c} \varphi_1^2 
+ w_{(A) 11} \hat{\varphi}_1 (-x Y_2 + e\bar{c} (\bar{\varphi}_1 + v)) 
    + w_{(A) 12} \bar{c} \hat{\varphi}_1^2
+ w_{(A) 13} \bar{c} \varphi_2^2 \right. \nonumber \\
& & \left. + w_{(A) 14} \hat{\varphi}_2 (-x Y_1 - e\bar{c} \bar{\varphi}_2 ) 
+ w_{(A)15} \bar{c} \hat{\varphi}_2^2 \right\}
\end{eqnarray}
All the coefficients $w_1, \dots , w_{15}, w_{A 1}, \dots , w_{A 15}$ in (\ref{gl6.18}) are of order
$\hbar$.\\
In the next step the consistency condition (\ref{gl6.17}) is used in order to determine the coefficients $w_i, w_{A i}$ as far as possible. Picking out in (\ref{gl6.17}) terms proportional to
$\chi$ we have
\begin{eqnarray}
\label{gl6.19}
  0 & = & \int \biggl( w_1 \bigl(\varphi_2 {\delta \hat{\Gamma}_{cl} 
  \over \delta \varphi_1  } 
  - Y_1 {\delta \hat{\Gamma}_{cl} \over \delta Y_2} \bigr) 
  + w_4 \bigl(\varphi_1 {\delta \hat{\Gamma}_{cl}  \over \delta \varphi_2  } 
  - Y_2 {\delta \hat{\Gamma}_{cl} \over \delta Y_1 }\bigr) 
  + w_3 {\delta \hat{\Gamma}_{cl}  \over \delta \varphi _2}  \\
  & & 
  + w_{14} \bigl(\hat \varphi_2 {\delta \hat{\Gamma}_{cl} 
  \over \delta \hat \varphi_1  }
  +q_2 {\delta \hat{\Gamma}_{cl} \over \delta q_1} \bigr) 
  + w_{11} \bigl(\hat \varphi_1 {\delta \hat{\Gamma}_{cl}  \over \delta \hat
  \varphi_2  } 
  +q_1 {\delta \hat{\Gamma}_{cl} \over \delta q_2 }\bigr) 
  + w_7 {\delta \hat{\Gamma}_{cl}  \over \delta \hat \varphi _2} \nonumber \\
  &&\,+  w_ 2\bigl(\hat \varphi_2 {\delta \hat{\Gamma}_{cl} 
  \over \delta \varphi_1  } 
  - Y_1 q_2 \bigr) 
  + w_5 \bigl(\hat \varphi_1 {\delta \hat{\Gamma}_{cl}  \over \delta \varphi_2  } 
  - Y_2 q_1 \bigr) \nonumber \\
  & & \,+ w_6 B + w_8 (B \hat \varphi _1 - \bar c q_1 ) + 
  w_{12}(B \hat \varphi _1^2
  -\bar c  s \hat \varphi _1^2 ) +
  w_{15}(B \hat \varphi _2^2
  -\bar c  s \hat \varphi _2^2 )  \nonumber \\
  && \,+ w_9 (B A^2 - \bar c sA^2 ) + 
  w_{10}(B  \varphi _1^2
  -\bar c  s  \varphi _1^2 ) +
  w_{13}(B  \varphi _2^2
  -\bar c  s  \varphi _2^2 ) \biggl)  \nonumber 
\end{eqnarray}
and a quite analogous expression for the terms proportional to $\chi_A$ obtained from (\ref{gl6.19}) by replacing all $w_i$ by the corresponding $w_{A i}$.\\
At this point we observe that the classical WI for $\chi = 0 = \chi_A$,
\begin{equation}
\label{gl6.20}
  \hat{W} \hat{\Gamma}_{cl} = 0 \; \; \; ,
\end{equation}
implies that  not all of the polynomials in (\ref{gl6.19}) can be independent. To proceed further, we hence have to eliminate one of these polynomials via (\ref{gl6.20}), for instance
$\varphi_2 \frac{\delta \hat{\Gamma}_{cl}}{\delta \varphi_1} -
Y_1 \frac{\delta \hat{\Gamma}_{cl}}{\delta Y_2}$. The remaining $14$ polynomials are independent and therefore their coefficients have to vanish due to (\ref{gl6.19}). In summary, algebraic considerations alone tell us that there possibly could exist a $\chi$- and $\chi_A$-anomaly
of the global WI (which nevertheless has to be $s_{\Gamma_{cl}}$-invariant because of
(\ref{gl6.17})):
\begin{eqnarray}
\label{gl6.21}
  (W^{gen} \Gamma )^{(\leq 1)} & = & \chi \Delta_{br}^{(\leq 1)} + \chi_A \Delta_{br_A}^{(\leq 1)} \\  
  & & \! \! \! \! + (\chi w_1 + \chi_A w_{A 1} ) \int \left(
  (Y_1 - e \bar{c} \hat{\varphi}_2 ) \bar{\varphi}_2 -
  (Y_2 + e \bar{c} (\hat{\varphi}_1 - \xi_A \frac{m}{e} ))
  (\bar{\varphi}_1 + \frac{m}{e} ) \right) \nonumber
\end{eqnarray}
Testing, however, (\ref{gl6.21}) with respect to $Y_1 \varphi_2$ and $Y_2 \varphi_1$ and making use of the fact that the 3-point functions disappear at an asymptotic momentum $p_{\infty}^2$,
\begin{eqnarray}
\label{gl6.22}
  \Gamma_{Y_1 \varphi_1}^{(1)} (p_{\infty}^2 ) + \Gamma_{Y_2 \varphi_2}^{(1)} (p_{\infty}^2 ) &
  = & - \chi w_1 - \chi_A w_{A 1} \\
  \Gamma_{Y_1 \varphi_1}^{(1)} (p_{\infty}^2 ) + \Gamma_{Y_2 \varphi_2}^{(1)} (p_{\infty}^2 ) &
  = & \chi w_1 + \chi_A w_{A 1} \; \; \; , \nonumber
\end{eqnarray}
we finally find:
\begin{equation}
\label{gl6.23}
  w_1 = 0 \; \; \; \mbox{ and } \; \; \; w_{A 1} = 0
\end{equation}
This concludes the proof of the $\chi$- and $\chi_A$-enlarged WI at 1-loop order.\\
It is clear that this result can immediately be generalized to all orders of the perturbative expansion by repeating the reasoning just given when proving the induction step: order $n$ in $\hbar \longrightarrow$ order $n+1$ in $\hbar$. Hence we have shown (\ref{gl6.10}) to all orders.

\newsection{The local Ward identity}

We conclude the first part of the present paper, which extends the results of \cite{HK} to the case
when all gauge parameters of the model undergo BRS transformations by looking at the local Ward identity. This local WI governs the invariance of Green functions under (deformed) local gauge transformations and also yields information about the $\xi$- and $\xi_A$-dependence of these Green functions. In analogy to the treatment of the global WI we again start with the local WI as it was proven in
\cite{KS} for $\chi = 0 =\chi_A$ to all orders of perturbation theory,
\begin{equation}
\label{gl7.1}
  \left. \left( (e + \delta e) w^{gen} (x) - \partial_{\mu} \frac{\delta}{\delta A_{\mu}} \right)
  \Gamma \right|_{\chi = 0 = \chi_A} = \Box B \; \; \; ,
\end{equation}
and then generalize to $\chi \neq 0$ and $\chi_A \neq 0$. In (\ref{gl7.1}) $w^{gen} (x)$ denotes the
($\chi$- and $\chi_A$-dependent) local Ward operator which is obtained from the global one
(\ref{gl6.7}) by taking away the integration,
\begin{equation}
\label{gl7.2}
  W^{gen} = \int d^4 x \; w^{gen} (x) \; \; \; ,
\end{equation}
and $\delta e$ -- to be fixed by the normalization condition for the coupling -- is of order $\hbar$.\\
In the course of proving the $\chi$- and $\chi_A$-dependent local WI it will turn out that the overall normalization factor of matter transformations $e + \delta e$ has to be independent of both
gauge parameters $\xi$ and $\xi_A$ to all orders in the loop expansion. This fact will be the most important result of the actual investigation. It exactly reflects the restrictions found in 
section 5 for the $\xi$- and $\xi_A$-dependence of the vertex $\Gamma_{A_{\mu} \varphi_1
\varphi_2}$ (\ref{gl5.10}) at the level of the local WI.\\
In the classical approximation a straightforward calculation shows that the following local WI holds true:
\begin{equation}
\label{gl7.3}
  (e w^{gen} (x) - \partial_{\mu} \frac{\delta}{\delta A_{\mu}} ) \Gamma_{cl}^{gen} =
  \Box B + e \chi D_{br} (x) + e \chi_A D_{br_A} (x)
\end{equation}
$D_{br} (x)$ and $D_{br_A} (x)$ are the (classical) non-integrated breaking terms
$\Delta_{br}$ and $\Delta_{br_A}$ (\ref{gl6.9}), respectively:
\begin{equation}
\label{gl7.4}
  \Delta_{br_{(A)}} = \int d^4 x \; D_{br_{(A)}} (x)
\end{equation}
In order to proceed to higher orders we will make use of the same two, general, ingredients which allowed us to prove the global WI, namely the action principle and the transformation behaviour of the local Ward operator $w^{gen} (x)$ (\ref{gl7.2}) under BRS transformations. Taking into account
the validity of the global WI (\ref{gl6.10}) and the local WI for $\chi = 0 = \chi_A$ the action principle implies that at 1-loop order we have:
\begin{eqnarray}
\label{gl7.5}
  & \left. \left( (e + \delta e^{(1)} ) w^{gen} (x) - \partial_{\mu} 
  \displaystyle\frac{\delta}{\delta A_{\mu}}
  \right) \Gamma \right|^{(\leq 1)} =
  \Box B + (e + \delta e^{(1)} ) \chi D_{br}^{(\leq 1)} (x) & \nonumber \\
  & + (e + \delta e^{(1)} ) \chi_A D_{br_A}^{(\leq 1)} (x)
  + \chi \partial^{\mu} j_{\mu}^{(1)} +
  \chi_A \partial^{\mu} j_{A \; \mu}^{(1)} +
  \chi \chi_A \partial^{\mu} j_{\chi \chi_A \; \mu}^{(1)} &
\end{eqnarray}
The currents $j_{\mu}^{(1)} (x), j_{A\; \mu}^{(1)} (x)$ and $j_{\chi \chi_A \; \mu}^{(1)} (x)$ have dimension less than or equal to three and are odd under charge conjugation. Furthermore,
$j_{\mu}^{(1)} (x)$ and $j_{A \; \mu}^{(1)} (x)$ carry $\phi \pi$-charge $-1$, whereas
$j_{\chi \chi_A \; \mu}^{(1)} (x)$ has $\phi \pi$-charge $-2$. Looking at the quantum numbers of the
fields in question we find that there is no possible term contributing to 
$j_{\chi \chi_A \; \mu}^{(1)} (x)$,
\begin{equation}
\label{gl7.6}
  j_{\chi \chi_A \; \mu}^{(1)} (x) \equiv 0 \; \; \; ,
\end{equation}
and only one contribution to $j_{\mu}^{(1)} (x)$ and $j_{A \; \mu}^{(1)} (x)$, respectively:
\begin{equation}
\label{gl7.7}
  j_{\mu}^{(1)} (x) = u \; \partial_{\mu} \bar{c} \; \; \mbox{ and } \; \;
  j_{A \; \mu}^{(1)} (x) = u_A \; \partial_{\mu} \bar{c}
\end{equation}
Next we have to exploit the transformation behaviour of $w^{gen} (x)$ (\ref{gl7.2}) under BRS
transformations; in direct analogy to the corresponding considerations (\ref{gl6.12}), (\ref{gl6.13}) for the global Ward operator we deduce
\begin{equation}
\label{gl7.8}
  0 = w^{gen} (x) {\cal S} (\Gamma ) =
  s_{\Gamma} \left( w^{gen} (x) \Gamma - \chi D_{br} (x) - \chi_A D_{br_A} (x) \right)
\end{equation}
and also:
\begin{equation}
\label{gl7.9}
  0 = \partial_{\mu} \frac{\delta}{\delta A_{\mu}} {\cal S} (\Gamma ) =
  s_{\Gamma} \left( \partial_{\mu} \frac{\delta}{\delta A_{\mu}} \Gamma \right)
\end{equation}
Therefore, acting with the $\chi$- and $\chi_A$-dependent $s_{\Gamma}$ on (\ref{gl7.5}) and making use of the consistency condition (\ref{gl7.8}) and (\ref{gl7.9}) one obtains after a short calculation the following algebraic constraint:
\begin{eqnarray}
\label{gl7.10}
  & \chi [\partial_{\xi} (e + \delta e^{(1)} ) ] (w^{gen} (x) \Gamma - \chi_A D_{br_A} (x) ) +
      \chi_A [\partial_{\xi_A} (e + \delta e^{(1)} ) ] (w^{gen} (x) \Gamma - \chi D_{br} (x) ) &
   \nonumber \\
   & = \chi \chi_A (\partial_{\xi} u_A - \partial_{\xi_A} u ) \Box \bar{c} -
      \chi u \Box B - \chi_A u_A \Box B &
\end{eqnarray}
Singling out in (\ref{gl7.10}) terms proportional to $\chi$ or $\chi_A$, respectively, we find at
1-loop order:
\begin{eqnarray}
\label{gl7.11}
  \chi & : & \partial_{\xi} \delta e^{(1)} \left. \left( w^{gen} (x) \Gamma 
  \right) \right|_{\chi = 0 = \chi_A}^{(0)} + u \Box B = 0 \\
  \chi_A & : & \partial_{\xi_A} \delta e^{(1)} \left. \left( w^{gen} (x) \Gamma \right)
  \right|_{\chi = 0 = \chi_A}^{(0)} + u_A \Box B = 0
\end{eqnarray}
Because the two insertions $\left. w^{gen} (x) \Gamma \right|_{\chi = 0 = \chi_A}^{(0)}$ and
$\Box B$ are linearly independent (as can be seen by directly calculating
$\left. w^{gen} (x) \Gamma \right|_{\chi = 0 = \chi_A}^{(0)}$, see also \cite{HK}) the coefficients in front of these two insertions have to vanish separately, i.e.:
\begin{equation}
\label{gl7.13}
  u = 0 = u_A \; \; \mbox{ and } \; \;
  \partial_{\xi} \delta e^{(1)} = 0 = \partial_{\xi_A} \delta e^{(1)}
\end{equation}
Inserting $u = 0 = u_A$ into (\ref{gl7.5}) completes the proof of the local WI at 1-loop order.\\
It is obvious that the argument just given can be extended to all orders of perturbation theory
by induction. Hence we finally end up with the following $\chi$- and $\chi_A$-dependent local
WI:
\begin{equation}
\label{gl7.14}
  \left( (e + \delta e ) w^{gen} (x) - \partial_{\mu} \frac{\delta}{\delta A_{\mu}} \right) \Gamma =
  \Box B + (e + \delta e ) \chi D_{br} (x) + (e + \delta e ) \chi_A D_{br_A} (x)
\end{equation}
In addition we have shown that the overall normalization factor of the matter transformations has to be $\xi$- and $\xi_A$-independent in all orders of the perturbative expansion:
\begin{equation}
\label{gl7.15}
  \partial_{\xi} (e + \delta e ) = 0 = \partial_{\xi_A} ( e + \delta e )
\end{equation}
This result is highly non-trivial and can be obtained in this generality only with the formalism
of BRS transforming gauge parameters.\\
As already mentioned above the constraint (\ref{gl7.15}) found at the level of the local WI is the
direct analogue of the restriction (\ref{gl5.10}) which we derived for the $\xi$- and 
$\xi_A$-dependence of the vertex $\Gamma_{A_{\mu} \varphi_1 \varphi_2}$. In section 5 we
also discussed that this restriction carefully has to be taken into account when a normalization
condition for the coupling is formulated, leading in section 5 to the introduction of two additional parameters into the theory, namely the two reference points $\xi_0$ and $\xi_{A_0}$, see
(\ref{gl5.11}). The normalization condition (\ref{gl5.11}), however, poses quite troublesome difficulties when explicit calculations are to be performed.\\
But, having proven (\ref{gl7.15}) we have at hand a new possibility for fixing the 
coupling\footnote{See also \cite{HK} for a more detailed discussion.}: Following the line of argument, the normalization condition for the coupling has to respect the $\xi$- and
$\xi_A$-independence of the factor $e + \delta e$. This is trivially fulfilled if we demand 
$\delta e = 0$, i.e. if we require the local WI to be exact to all orders of perturbation theory:
\begin{equation}
\label{gl7.16}
  \left. \left( e w^{gen} (x) - \partial_{\mu} \frac{\delta}{\delta A_{\mu}} \right) \Gamma
  \right|_{\chi = 0 = \chi_A} = \Box B
\end{equation}
The normalization condition (\ref{gl7.16}) (replacing (\ref{gl5.11})) is much easier manageable in concrete calculations.\\
In summary, we have shown that the on-shell normalization conditions taken together with the
requirement ``local WI exact to all orders'' are in agreement with the $\chi$- and 
$\chi_A$-enlarged ST identity and hence guarantuee a correct treatment of full gauge parameter dependence in explicit calculations.

\newsection{BRS-symmetric insertions}

As an application of the general formalism developed so far we want to study parametric
differential equations of the type
\begin{equation}
\label{gl8.1}
  \lambda \partial_{\lambda} \Gamma = \Delta_{\lambda} \cdot \Gamma
\end{equation}
in the next section, where $\lambda$ denotes a (set of) parameter(s) of the theory.
Due to the action principle $\Delta_{\lambda}$ is an insertion of dimension less than or
equal to four, even under charge conjugation and BRS invariant. This last property holds
because of
\begin{equation}
\label{gl8.2}
  0 = \lambda \partial_{\lambda} {\cal S} (\Gamma ) 
     = s_{\Gamma} (\lambda \partial_{\lambda} \Gamma )
     = s_{\Gamma} (\Delta_{\lambda} \cdot \Gamma )
     = s_{\Gamma_{cl}} \Delta_{\lambda} + {\cal O} (\hbar )
\end{equation}
for $\lambda$ being independent of $\xi$ and $\xi_A$. Therefore, as a preparatory step we first
have to classify all BRS-symmetric insertions, which carry the same quantum numbers as
$\Gamma$. Because in the present paper we are mainly interested in questions concerning
gauge parameter dependence we will pay special attention to the appearing of $\xi$- and
$\xi_A$-dependence.\\
In order to solve the cohomological problem mentioned above we once more return to the
classical level and write down all independent field polynomials fulfilling
\begin{equation}
\label{gl8.3}
  s_{\Gamma_{cl}} \Delta_{\lambda} = 0 \; \; \; .
\end{equation}
Then we have to translate these polynomials to BRS-invariant operators, only this last
representation being valid to all orders of perturbation theory. Because the solution of this
problem for $\chi = 0 = \chi_A$ was already given in \cite{KS} we will make use of the following trick
to handle the modifications for $\chi \neq 0$, $\chi_A \neq 0$:\\
First we decompose $\Delta_{\lambda}$ into three parts by explicitly separating $\chi$- and
$\chi_A$-dependence:
\begin{equation}
\label{gl8.4}
  \Delta_{\lambda} = \Delta_{\lambda}^0 + \chi \Delta_{\lambda}^- + \chi_A \Delta_{A, \lambda}^-
\end{equation}
(Please note that due to the quantum numbers of $\Delta_{\lambda}$ no term proportional to
$\chi \chi_A$ can appear.)\\
Splitting $\Gamma_{cl} = \hat{\Gamma}_{cl} + \chi Q + \chi_A Q_A$ and $s_{\Gamma_{cl}}$ in the same way,
\begin{eqnarray}
\label{gl8.5}
  s_{\Gamma_{cl}} & = & s_{\hat{\Gamma}_{cl}}^{\chi = 0 = \chi_A} +
                                         \chi (\partial_{\xi} + {\cal O} ) +
					  \chi_A (\partial_{\xi_A} + {\cal O}_A ) \; \; \; , \\ 
  \mbox{with} \; \; \; \; \; {\cal O} & = & \int \left\{ \frac{\delta Q}{\delta \underline{\varphi}}
                                                           \frac{\delta}{\delta \underline{Y}} -
					       \frac{\delta Q}{\delta \underline{Y}} 
					       \frac{\delta}{\delta \underline{\varphi}} \right\} \; \; \; , \nonumber \\
  {\cal O}_A & = & \int \left\{ \frac{\delta Q_A}{\delta \underline{\varphi}}
                                             \frac{\delta}{\delta \underline{Y}} -
			       \frac{\delta Q_A}{\delta \underline{Y}}
			       \frac{\delta}{\delta \underline{\varphi}} \right\} \; \; \; , \nonumber
\end{eqnarray}
we find that (\ref{gl8.3}) is equivalent to the following four equations:
\begin{eqnarray}
\label{gl8.6}
  s_{\hat{\Gamma}_{cl}}^{\chi = 0 = \chi_A} \Delta_{\lambda}^0 & = & 0 \\
\label{gl8.7}
  s_{\hat{\Gamma}_{cl}}^{\chi = 0 = \chi_A} \Delta_{\lambda}^- & = &
    (\partial_{\xi} + {\cal O} ) \Delta_{\lambda}^0 \\
\label{gl8.8}
  s_{\hat{\Gamma}_{cl}}^{\chi = 0 = \chi_A} \Delta_{A, \lambda}^- & = &
    (\partial_{\xi_A} + {\cal O}_A ) \Delta_{\lambda}^0 \\
\label{gl8.9}
  (\partial_{\xi} + {\cal O} ) \Delta_{A, \lambda}^- & = &
    (\partial_{\xi_A} + {\cal O}_A ) \Delta_{\lambda}^-
\end{eqnarray}
Now it is easy to see that it is always possible to find a $\hat{\Delta}_{\lambda}^-$ such that:
\begin{equation}
\label{gl8.10}
  \Delta_{\lambda}^- = (\partial_{\xi} + {\cal O} ) \hat{\Delta}_{\lambda}^- \; \; \; \mbox{ and } \; \; \;
  \Delta_{A, \lambda}^- = (\partial_{\xi_A} + {\cal O}_A ) \hat{\Delta}_{\lambda}^-
\end{equation}
We remark that due to (\ref{gl8.10}) eq. (\ref{gl8.9}) is fulfilled automatically.\\
With these preparations we have
\begin{eqnarray}
\label{gl8.11}
  (\partial_{\xi} + {\cal O} ) (\Delta_{\lambda}^0 -
                                           s_{\hat{\Gamma}_{cl}}^{\chi = 0 = \chi_A} \hat{\Delta}_{\lambda}^- ) 
  & = & s_{\hat{\Gamma}_{cl}}^{\chi = 0 = \chi_A} \Delta_{\lambda}^- -
            (\partial_{\xi} + {\cal O} ) s_{\hat{\Gamma}_{cl}}^{\chi = 0 = \chi_A} 
	                                            \hat{\Delta}_{\lambda}^- \nonumber \\
  & = & s_{\hat{\Gamma}_{cl}}^{\chi = 0 = \chi_A} \Delta_{\lambda}^- -
            s_{\hat{\Gamma}_{cl}}^{\chi = 0 = \chi_A} (\partial_{\xi} + {\cal O} )
	                                            \hat{\Delta}_{\lambda}^- \; \; = \; \; 0
\end{eqnarray}
and in the same way
\begin{equation}
\label{gl8.12}
  (\partial_{\xi_A} + {\cal O}_A ) (\Delta_{\lambda}^0 -
                                                    s_{\hat{\Gamma}_{cl}}^{\chi = 0 = \chi_A} 
						      \hat{\Delta}_{\lambda}^- )  \; \; = \; \; 0 \; \; \; .
\end{equation}
But that means
\begin{equation}
\label{gl8.13}
  \Delta_{\lambda}^0 = s_{\hat{\Gamma}_{cl}}^{\chi = 0 = \chi_A} \hat{\Delta}_{\lambda}^-   +
                                    \hat{\Delta}_{\lambda}^0
\end{equation}
with
\begin{equation}
\label{gl8.14}
  (\partial_{\xi} + {\cal O} ) \hat{\Delta}_{\lambda}^0 = 0 =
  (\partial_{\xi_A} +{\cal O}_A ) \hat{\Delta}_{\lambda}^0
\end{equation}
and
\begin{equation}
\label{gl8.15}
  s_{\hat{\Gamma}_{cl}}^{\chi = 0 = \chi_A} \hat{\Delta}_{\lambda}^0 = 0 \; \; \; .
\end{equation}
Hence:
\begin{eqnarray}
\label{gl8.16}
  \Delta_{\lambda} & = & \Delta_{\lambda}^0 + \chi \Delta_{\lambda}^- +
                                        \chi_A \Delta_{A, \lambda}^- \nonumber \\
  & = & s_{\hat{\Gamma}_{cl}}^{\chi = 0 = \chi_A} \hat{\Delta}_{\lambda}^- +
            \hat{\Delta}_{\lambda}^0 +
	    \chi (\partial_{\xi} + {\cal O} ) \hat{\Delta}_{\lambda}^- +
	    \chi_A (\partial_{\xi_A} + {\cal O}_A ) \hat{\Delta}_{\lambda}^- \nonumber \\
  & = & \hat{\Delta}_{\lambda}^0 + s_{\Gamma_{cl}} \hat{\Delta}_{\lambda}^-
\end{eqnarray} 
As already mentioned, the solution of (\ref{gl8.15}) was presented in \cite{KS} and we just give
the list of terms contributing to $\hat{\Delta}_{\lambda}^0$ in appendix B. The crucial point in this context is, however: A short calculation starting from (\ref{gl8.14}) shows that all terms in $\hat{\Delta}_{\lambda}^0$ which are {\it no} BRS variations, namely\footnote{The definition of the additional external field $\hat{\varphi}_0$ is also given in appendix B.},
\begin{equation}
\label{gl8.17}
  \int \frac{\delta \hat{\Gamma}_{cl}}{\delta \hat{\varphi}_0} \; ,\;
  \int \left\{ A \frac{\delta}{\delta A} + c \frac{\delta}{\delta c} \right\} \hat{\Gamma}_{cl} \; ,\;
  m_H \partial_{m_H} \hat{\Gamma}_{cl} \; , \;
  e \partial_e \hat{\Gamma}_{cl} \; ,
\end{equation}
have to appear with coefficients which are {\it independent of} $\xi$ {\it and} $\xi_A$.\\
For $\hat{\Delta}_{\lambda}^-$ we choose the most general ansatz compatible with the quantum
numbers of $\hat{\Delta}_{\lambda}^-$ ($\phi \pi$-charge: -1, $C$: +, dim: $\leq 4$),
see also (\ref{gl4.8}); in view
of the generalization to higher orders, this ansatz can be brought into the form:
\begin{eqnarray}
\label{gl8.18}
  \hat{\Delta}_{\lambda}^- & : & \int Y_1, \frac{\delta \hat{\Gamma}_{cl}}{\delta q_1},
                                                       \bar{c} \hat{\varphi}_2,  \\
  & & Y_1 \varphi_1, Y_1 \hat{\varphi}_1, Y_2 \varphi_2, Y_2 \hat{\varphi}_2,
         \bar{c} \frac{\delta \hat{\Gamma}_{cl}}{\delta B},
	 \hat{\varphi}_i \frac{\delta \hat{\Gamma}_{cl}}{\delta q_i},
	 \bar{c} \hat{\varphi}_1 \hat{\varphi}_2, \bar{c} B \nonumber 
\end{eqnarray}
Looking to (\ref{glB.1}) -- (\ref{glB.4}) and (\ref{gl8.16}), we find that all terms in $\hat{\Delta}_{\lambda}^0$ which are BRS variations have to be modified when BRS transforming
gauge parameters are included. Therefore we finally end up with the following basis of BRS
invariant insertions which we directly give in the form of BRS invariant operators ($i = 1,2$):
\begin{eqnarray}
\label{gl8.19}
  & & f_{s,0} \displaystyle\int \displaystyle\frac{\delta \Gamma}{\delta \varphi_1} +
     (\chi \partial_{\xi} + \chi_A \partial_{\xi_A} ) f_{s,0} \int Y_1 \; \; = \; \;
     s_{\Gamma} (f_{s,0} \displaystyle\int Y_1 ) \nonumber \\
  & & \hat{f}_{s,0} \displaystyle\int \displaystyle\frac{\delta \Gamma}{\delta \hat{\varphi}_1} +
     [(\chi \partial_{\xi} + \chi_A \partial_{\xi_A} ) \hat{f}_{s,0} ] \displaystyle\int
     \displaystyle\frac{\delta \Gamma}{\delta q_1} \; \; = \; \;
     s_{\Gamma} (\hat{f}_{s,0} \displaystyle\int \frac{\delta \Gamma}{\delta q_1} ) \nonumber \\
  & & \tilde{f}_3 \int \left\{ B \hat{\varphi}_2 - \bar{c} q_2 \right\} +
     (\chi \partial_{\xi} + \chi_A \partial_{\xi_A} ) \tilde{f}_3 \int \bar{c} \hat{\varphi}_2\; \; = \; \;
     s_{\Gamma} (\tilde{f}_3 \int \bar{c} \hat{\varphi}_2 ) \\
\label{gl8.20}
  {\cal N}^{(\chi, \chi_A)}_{s,i} \Gamma & \equiv
  &  f_{s,i} \displaystyle\int \left\{ \varphi_i \displaystyle\frac{\delta}{\delta \varphi_i} - 
      Y_i \displaystyle\frac{\delta}{\delta Y_i} \right\} \Gamma
     + (\chi \partial_{\xi} + \chi_A \partial_{\xi_A} ) f_{s,i} 
     \displaystyle\int Y_i \varphi_i \;\; = \; \; 
     s_{\Gamma} (f_{s,i} \displaystyle\int Y_i \varphi_i ) \nonumber \\
  \hat{\cal N}^{(\chi, \chi_A)}_{s,i} \Gamma & \equiv
  & \hat{f}_{s,i} \displaystyle\int \left\{ q_i \displaystyle\frac{\delta}{\delta q_i} + 
     \hat{\varphi}_i \displaystyle\frac{\delta}{\delta \hat{\varphi}_i} \right\} \Gamma
    + [(\chi \partial_{\xi} + \chi_A \partial_{\xi_A} ) \hat{f}_{s,i} ]
     \displaystyle\int \hat{\varphi}_i \displaystyle\frac{\delta \Gamma}{\delta q_i} \; \; = \; \;
     s_{\Gamma} (\hat{f}_{s,i} \displaystyle\int 
     \hat{\varphi}_i \frac{\delta \Gamma}{\delta q_i} ) \nonumber \\
  \bar{\cal N}^{(\chi, \chi_A)}_{s,i} \Gamma & \equiv
  & \bar{f}_{s,i} \displaystyle\int \left\{ 
      \hat{\varphi}_i \displaystyle\frac{\delta \Gamma}{\delta \varphi_i} - Y_i q_i \right\}
     + (\chi \partial_{\xi} + \chi_A \partial_{\xi_A} ) 
     \bar{f}_{s,i} \displaystyle\int Y_i \hat{\varphi}_i \; \; = \; \;
     s_{\Gamma} (\bar{f}_{s,i} \displaystyle\int Y_i \hat{\varphi}_i ) \nonumber \\
  {\cal N}^{(\chi, \chi_A)}_B \Gamma & \equiv
  & f_B \displaystyle\int \left\{ B \displaystyle\frac{\delta}{\delta B} + 
      \bar{c} \displaystyle\frac{\delta}{\delta \bar{c}} \right\} \Gamma
     + [(\chi \partial_{\xi} + \chi_A \partial_{\xi_A} ) f_B ] 
     \displaystyle\int \bar{c} \displaystyle\frac{\delta \Gamma}{\delta B} \; \; = \; \;
     s_{\Gamma} (f_B \displaystyle\int \bar{c} \frac{\delta \Gamma}{\delta B} ) \nonumber \\
  & & \! \! \! \! \! \! \! \! \! \! \tilde{f}_4
                      \int \left\{ B \hat{\varphi}_1 \hat{\varphi}_2 - 
                      \bar{c} q_1 \hat{\varphi}_2 -
                      \bar{c} \hat{\varphi}_1 q_2 \right\}
     + (\chi \partial_{\xi} + \chi_A \partial_{\xi_A} ) \tilde{f}_4
     \int \bar{c} \hat{\varphi}_1 \hat{\varphi}_2 \; \; = \; \;
     s_{\Gamma} (\tilde{f}_4 \int \bar{c} \hat{\varphi}_1 \hat{\varphi}_2 ) \nonumber \\
  & & \! \! \! \! \! \! \! \! \! \! f_{\xi} \partial_{\xi} \Gamma +
     [(\chi \partial_{\xi} + \chi_A \partial_{\xi_A} ) f_{\xi} ] \partial_{\chi} \Gamma \; \; = \; \;
     s_{\Gamma} (f_{\xi} \partial_{\chi} \Gamma )
\end{eqnarray}
In addition there are the BRS symmetric operators from (\ref{gl8.17}) (with $\hat{\Gamma}_{cl}$
replaced by $\Gamma$).

\newsection{Parametric differential equations}

Having finished the preparatory considerations dealing with the BRS symmetric insertions we now can turn to the derivation of some partial differential equations, namely the Callan-Symanzik (CS) equation and the renormalization group (RG) equation. We will also comment about the dependence of the theory on the ghost mass which is governed by the differential operator $\xi_A \partial_{\xi_A}$ due to the normalization condition (\ref{massnorm}).

\newsubsection{CS equation}

The CS equation describes the response of the system to the scaling of all {\it independent} parameters carrying dimension of mass. In the model under investigation the CS operator is hence given by
\begin{equation}
\label{gl9.1}
  \underline{m} \partial_{\underline{m}} \equiv m \partial_m + m_H \partial_{m_H} +
  \kappa \partial_{\kappa}
\end{equation}
and we have the task to construct the r.h.s. of $\underline{m} \partial_{\underline{m}} \Gamma
= \; \; ?$ which according to the action principle
\begin{equation}
\label{gl9.2}
  \underline{m} \partial_{\underline{m}} \Gamma = \Delta_m \cdot \Gamma
\end{equation}
has to be an insertion with dimension less than or equal to four, even under charge conjugation and also BRS invariant. (I.e. $\underline{m} \partial_{\underline{m}}$ is an operator of the type 
$\lambda \partial_{\lambda}$ discussed in the previous section.) In \cite{KS} it was shown that in order to construct a {\it unique} r.h.s. of the CS equation rigid invariance has to be used, too. Therefore we next calculate the commutator of the $\chi$- and $\chi_A$-enlarged global Ward operator (\ref{gl6.7}) and $\underline{m} \partial_{\underline{m}}$:
\begin{equation}
\label{gl9.3}
  [ W^{gen} , \underline{m} \partial_{\underline{m}} ] =
  z \int \left\{ \hat{\xi}_A \frac{m}{e} \frac{\delta}{\delta \varphi_2} +
  \xi_A \frac{m}{e} \frac{\delta}{\delta \hat{\varphi}_2} \right\} +
  (\chi \partial_{\xi} + \chi_A \partial_{\xi_A} ) z \xi_A \frac{m}{e}
  \int \frac{\delta}{\delta q_2}
\end{equation}
In order to make the line of argument as transparent as possible and to explicitly work out what is needed in the following we introduce the $W^{gen}$-symmetric extension $\underline{m}
\tilde{\partial}_{\underline{m}}$ of $\underline{m} \partial_{\underline{m}}$ originating from
(\ref{gl9.3}),
\begin{eqnarray}
\label{gl9.4}
  & \underline{m} \tilde{\partial}_{\underline{m}} \equiv \underline{m} \partial_{\underline{m}} +
  \hat{\xi}_A \displaystyle\frac{m}{e} \displaystyle\int \displaystyle\frac{\delta}{\delta \varphi_1} +
  \xi_A \frac{m}{e} \displaystyle\int \displaystyle\frac{\delta}{\delta \hat{\varphi}_1} +
  (\chi \partial_{\xi} + \chi_A \partial_{\xi_A} ) \xi_A \frac{m}{e} 
  \displaystyle\int \displaystyle\frac{\delta}{\delta q_1}
  \; \; \; , & \\
\label{gl9.5}
  & [ W^{gen} , \underline{m} \tilde{\partial}_{\underline{m}} ] = 0 \; \; \; , &
\end{eqnarray}
and consider the insertion $\underline{m} \tilde{\partial}_{\underline{m}} \Gamma =
\tilde{\Delta}_m \cdot \Gamma$ instead of $\underline{m} \partial_{\underline{m}} \Gamma =
\Delta_m \cdot \Gamma$. Due to (\ref{gl9.5}) we have:
\begin{equation}
\label{gl9.6}
  W^{gen} (\tilde{\Delta}_m \cdot \Gamma ) = 
  \underline{m} \tilde{\partial}_{\underline{m}} W^{gen} \Gamma =
  \underline{m} \tilde{\partial}_{\underline{m}} (\chi \Delta_{br} + \chi_A \Delta_{br_A} ) =
  z (\chi \partial_{\xi} + \chi_A \partial_{\xi_A} ) \hat{\xi}_A \frac{m}{e} \int Y_2
\end{equation}
Please note that the application of $W^{gen}$ to the term
\begin{equation}
\label{gl9.7}
  - (\chi \partial_{\xi} + \chi_A \partial_{\xi_A} ) \hat{\xi}_A \frac{m}{e} \int Y_1 \; \; \; ,
\end{equation}
this term being part of the first insertion in (\ref{gl8.19}), exactly cancels the r.h.s. of (\ref{gl9.6}). Therefore, all other BRS symmetric\footnote{The operators extending $\underline{m}
\partial_{\underline{m}}$ in (\ref{gl9.4}) taken together with (\ref{gl9.7}) (times $-1$) just constitute the first two BRS symmetric insertions in (\ref{gl8.19}); hence the remaining contributions to $\tilde{\Delta}_m \cdot \Gamma$ have to be BRS symmetric.} insertions building up 
$\tilde{\Delta}_m \cdot \Gamma$ have to be symmetrized with respect to $W^{gen}$: Only in this
$W^{gen}$-symmetrized form they can contribute to the r.h.s. of the CS equation. For some of the operators in (\ref{gl8.17}), (\ref{gl8.19}), (\ref{gl8.20}) (namely the ($\chi$- and $\chi_A$-enlarged) leg counting operators) this symmetrization can be achieved easily:
\begin{eqnarray}
\label{gl9.8}
  {\cal N}_s^{(\chi , \chi_A )} \Gamma & \equiv &
  f_s N_s \Gamma - f_s \hat{\xi}_A \frac{m}{e} \int \frac{\delta \Gamma}{\delta \varphi_1} +
  (\chi \partial_{\xi} + \chi_A \partial_{\xi_A} ) f_s \int \left\{
  Y_1 (\varphi_1 - \hat{\xi}_A \frac{m}{e} ) + Y_2 \varphi_2 \right\} \; \; \; , \nonumber \\
  \hat{\cal N}_s^{(\chi , \chi_A )} & \equiv &
  \hat{f}_s \hat{N}_s - \hat{f}_s \xi_A \frac{m}{e} \int \frac{\delta}{\delta \hat{\varphi}_1} +
  (\chi \partial_{\xi} + \chi_A \partial_{\xi_A} ) \hat{f}_s \int \left\{
  (\hat{\varphi}_1 - \xi_A \frac{m}{e} ) \frac{\delta}{\delta q_1} +
  \hat{\varphi}_2 \frac{\delta}{\delta q_2} \right\} \; \; \; , \nonumber \\
  N_A & \equiv &
  \int \left\{ A \frac{\delta}{\delta A} + c \frac{\delta}{\delta c} \right\} \; \; \; , \nonumber \\
  {\cal N}_B^{(\chi , \chi_A )} & \equiv &
  f_B N_B + 
  (\chi \partial_{\xi} + \chi_A \partial_{\xi_A} ) f_B \int \bar{c} \frac{\delta}{\delta B}
\end{eqnarray}
The mixed operators containing $\hat{\varphi}_i \frac{\delta \Gamma}{\delta \varphi_i}$ are symmetrized like the leg counting operators:
\begin{eqnarray}
\label{gl9.9}
  \bar{\cal N}_s^{(\chi , \chi_A )} \Gamma & \equiv &
  \bar{f}_s \bar{N}_s \Gamma - 
  \bar{f}_s \xi_A \frac{m}{e} \int \frac{\delta \Gamma}{\delta \varphi_1} +
  \bar{f}_s \int \left\{ q_1 Y_1 + q_2 Y_2 \right\} \nonumber \\
  & & + (\chi \partial_{\xi} + \chi_A \partial_{\xi_A} ) \bar{f}_s \int \left\{
  Y_1 (\hat{\varphi}_1 - \xi_A \frac{m}{e} ) + Y_2 \hat{\varphi}_2 \right\}
\end{eqnarray}
In (\ref{gl9.8}), (\ref{gl9.9}) we have introduced the usual leg counting operators:
\begin{eqnarray}
\label{gl9.9a}
  N_s & \equiv & \int \left\{ \varphi_1 \frac{\delta}{\delta \varphi_1} +
                                          \varphi_2 \frac{\delta}{\delta \varphi_2} -
  Y_1 \frac{\delta}{\delta Y_1} - Y_2 \frac{\delta}{\delta Y_2} \right\} \; \; \; , \nonumber \\
  \hat{N}_s & \equiv & \int \left\{ \hat{\varphi}_1 \frac{\delta}{\delta \hat{\varphi}_1} +
                                                  \hat{\varphi}_2 \frac{\delta}{\delta \hat{\varphi}_2} +
  q_1 \frac{\delta}{\delta q_1} + q_2 \frac{\delta}{\delta q_2} \right\} \; \; \; , \nonumber \\
  \bar{N}_s & \equiv & \int \left\{ \hat{\varphi}_1 \frac{\delta}{\delta \varphi_1} +
                                                   \hat{\varphi}_2 \frac{\delta}{\delta \varphi_2} \right\}
						   \; \; \; , \nonumber \\
  N_B & \equiv & \int \left\{ B \frac{\delta}{\delta B} +
                                          \bar{c} \frac{\delta}{\delta \bar{c}} \right\}
\end{eqnarray}
To find the $W^{gen}$-symmetric extensions of the differential operators $m_H \partial_{m_H}$,
$e \partial_e$ and the operator containing $\partial_{\xi}$ (see last line of (\ref{gl8.20})),
\begin{equation}
\label{gl9.10}
  m_H \partial_{m_H} \rightarrow m_H \tilde{\partial}_{m_H} \; \; , \; \;
  e \partial_e \rightarrow e \tilde{\partial}_e \; \; , \; \;
  f_{\xi} \partial_{\xi} + (\chi \partial_{\xi} + \chi_A \partial_{\xi_A} ) f_{\xi} \partial_{\chi}
  \rightarrow \tilde{\partial}_{\xi} \; \; \; ,
\end{equation}
indeed requires some calculation. The final expressions being rather lengthy we present the explicit results of the symmetrization in appendix C.\\
Finally we observe that the insertion $\frac{\delta}{\delta \hat{\varphi}_0}$ already is
$W^{gen}$-symmetric and that the remaining two insertions in (\ref{gl8.19}), (\ref{gl8.20}) cannot be extended in a $W^{gen}$-symmetric way.\\
Thus the final answer is: (\ref{gl9.8}), (\ref{gl9.9}), (\ref{glC.5}) and $\frac{\delta}{\delta \hat{\varphi}_0}$ provide a basis of BRS symmetric and rigidly invariant operators which are even under charge conjugation and have dimension less than or equal to four. Hence the insertion
$\tilde{\Delta}_m \cdot \Gamma$ can be decomposed as follows:
\begin{eqnarray}
\label{gl9.11}
  \widetilde{\cal C} \Gamma & \equiv &
  \left( \underline{m} \tilde{\partial}_{\underline{m}} + \beta_e e \tilde{\partial}_e +
  \beta_{m_H} m_H \tilde{\partial}_{m_H} + \tilde{\beta}_{\xi} \tilde{\partial}_{\xi} \right. \nonumber \\
  & & \left. - \tilde{\gamma}_s {\cal N}_s^{(\chi , \chi_A )} -
  \tilde{\hat{\gamma}}_s \hat{\cal N}_s^{(\chi , \chi_A )} -
  \tilde{\bar{\gamma}}_s \bar{\cal N}_s^{(\chi , \chi_A )} -
  \gamma_A N_A - \tilde{\gamma}_B {\cal N}_B^{(\chi , \chi_A )} -
  \alpha_{inv} \int \delta_{\hat{\varphi}_0} \right) \Gamma \nonumber \\
  & = & - (\chi \partial_{\xi} + \chi_A \partial_{\xi_A} ) \hat{\xi}_A \frac{m}{e} \int Y_1
\end{eqnarray}
Equation (\ref{gl9.11}) is the CS equation in the manifestly $W^{gen}$-symmetric form. The important result in our context is that the $\beta$-functions $\beta_e$ and $\beta_{m_H}$ as well
as the anomalous dimension $\gamma_A$ and $\alpha_{inv}$ 
are independent of both the gauge parameters
$\xi$ and $\xi_A$ to all orders of perturbation theory. The coefficient functions $\tilde{\beta}_{\xi}, \tilde{\gamma}_s, \tilde{\hat{\gamma}}_s, \tilde{\bar{\gamma}}_s, \tilde{\gamma}_B$ also are $\xi$- and $\xi_A$-independent, but the usual (i.e. complete) $\beta$- and $\gamma$-functions
\begin{equation}
\label{gl9.11a}
  \beta_{\xi} \xi = \tilde{\beta}_{\xi} f_{\xi} \; , \;
  \gamma_s = \tilde{\gamma}_s f_s \; , \;
  \hat{\gamma}_s = \tilde{\hat{\gamma}}_s \hat{f}_s \; ,\;
  \bar{\gamma}_s = \tilde{\bar{\gamma}}_s \bar{f}_s \; , \;
  \gamma_B = \tilde{\gamma}_B f_B
\end{equation}
a priori may depend on both the gauge parameters $\xi$ and $\xi_A$ through the factors
$f_{(s, \xi )} (\xi, \xi_A )$ appearing in the leg counting
operators ${\cal N}_s^{(\chi , \chi_A )}$ (\ref{gl9.8}), (\ref{gl9.9}) and in $\tilde{\partial}_{\xi}$ (\ref{glC.5}).\\
So far one can get with symmetry considerations alone. If additional information about the coefficient functions is requested one has to test (\ref{gl9.11}) on the gauge condition (\ref{gl4.2}), to make use of the local WI (\ref{gl7.14}) and/or to carry out explicit calculations: Testing (\ref{gl9.11}) on the gauge condition (\ref{gl4.2}) we find
\begin{eqnarray}
\label{gl9.11b}
  \gamma_B & = & - \gamma_A \\
  \beta_{\xi} \; \; = \; \; 2 \gamma_B & = & - 2 \gamma_A \nonumber \\
  \beta_e + \gamma_A - \gamma_s - \hat{\gamma}_s & = &
  (\beta_e e \partial_e + \beta_{m_H} m_H \partial_{m_H} -
   2 \gamma_A \xi \partial_{\xi} ) \mbox{ln} z \nonumber
\end{eqnarray}
and hence also $\gamma_B$ and $\beta_{\xi}$ are completely gauge parameter-independent to all orders. Furthermore, using the validity of the local WI (\ref{gl7.14}) and the normalization condition
for the coupling (\ref{gl7.16}) yields (see \cite{KS} for details):
\begin{equation}
\label{gl9.13a}
  \gamma_A = \beta_e
\end{equation}
We want to conclude this subsection by rewriting the CS equation in its much more convenient form which separates the hard and soft breaking on the left and right hand side of the CS
equation:
\begin{eqnarray}
\label{gl9.13b}
  & & \left( \underline{m} \partial_{\underline{m}} + \beta_e e \partial_e + 
                                                       \beta_{m_H} m_H \partial_{m_H} -
  \gamma_s N_s - \hat{\gamma}_s \hat{N}_s - \bar{\gamma}_s \bar{N}_s -
  \beta_e ( N_A - N_B + 2 \xi \partial_{\xi} ) \right. \nonumber \\
  & & \left. - \gamma_1 \int \left\{ \varphi_1 \frac{\delta}{\delta \varphi_1} - 
                                                  Y_1 \frac{\delta}{\delta Y_1} +
                                           \hat{\varphi}_1 \frac{\delta}{\delta \hat{\varphi}_1} +
					   q_1 \frac{\delta}{\delta q_1} \right \} \right. \\
  & & \left. - (\chi \partial_{\xi} + \chi_A \partial_{\xi_A} ) \int \left\{
         - \gamma_s \hat{\varphi}_1 \frac{\delta}{\delta q_1} +
	 \hat{\gamma}_s \hat{\varphi}_2 \frac{\delta}{\delta q_2} \right\} -
	 2 \beta_e \chi \partial_{\chi} \right) \Gamma \nonumber \\
  & = &  - \frac{m}{e} \int \left\{
            (\hat{\xi}_A + \alpha_1 ) \frac{\delta}{\delta \varphi_1} +
	    (\xi_A + \hat{\alpha}_1 ) \frac{\delta}{\delta \hat{\varphi}_1} -
	    \alpha_{inv} \frac{e}{m} \frac{\delta}{\delta \hat{\varphi}_0} +
	    (\chi \partial_{\xi} + \chi_A \partial_{\xi_A} ) (\xi_A + \hat{\alpha}_1 )
	    \frac{\delta}{\delta q_1} \right \} \Gamma \nonumber \\
  &  & + \bar{\gamma}_s \int \left\{ q_1 Y_1 + q_2 Y_2 \right\} \nonumber \\
  &  & + (\chi \partial_{\xi} + \chi_A \partial_{\xi_A} ) \int \left\{
	 (\gamma_s + \gamma_1 ) Y_1 \varphi_1 + \gamma_s Y_2 \varphi_2 -
	 \frac{m}{e} (\hat{\xi}_A + \alpha_1 ) Y_1 + \bar{\gamma}_s (Y_1 \hat{\varphi}_1 +
	 Y_2 \hat{\varphi}_2 ) \right\} \nonumber
\end{eqnarray}
with
\begin{eqnarray}
\label{gl9.13c}
  \gamma_1 & = & (\beta_e e \partial_e + \beta_{m_H} m_H \partial_{m_H} -
                               2 \beta_e \xi \partial_{\xi} ) \mbox{ln} z \; \; \; = 
			       \; \; \; {\cal O} (\hbar^2 ) \nonumber \\
  \hat{\gamma}_s & = & 2 \beta_e - \gamma_s - \gamma_1 \nonumber \\
  \alpha_1 & = & (\gamma_1 + \gamma_s - \beta_e ) \hat{\xi}_A + \bar{\gamma}_s \xi_A +
                           (\beta_e e \partial_e + \beta_{m_H} m_H \partial_{m_H} -
                               2 \beta_e \xi \partial_{\xi} ) \hat{\xi}_A \nonumber \\
  \hat{\alpha}_1 & = & (\gamma_1 + \hat{\gamma}_s - \beta_e ) \xi_A
\end{eqnarray}
In (\ref{gl9.13b}), (\ref{gl9.13c}) we have already incorporated the relations (\ref{gl9.11b}) and
(\ref{gl9.13a}). Therefore only the coefficient functions $\beta_e, \beta_{m_H}, \gamma_s,
\bar{\gamma}_s$ and the coefficient $\alpha_{inv} = \frac{1}{2} m_H^2 + {\cal O} (\hbar )$ of the soft insertion $\int \delta_{\hat{\varphi}_0}$ turn out to be independent and have to be determined by explicit calculations (see \cite{KS}).

\newsubsection{Dependence on the ghost mass}

Due to the normalization condition (\ref{massnorm}) the dependence of the theory on the ghost mass is encoded in the differential operator $\xi_A \partial_{\xi_A}$ and we have to analyse the r.h.s. of $\xi_A \partial_{\xi_A} \Gamma = \; \; ?$. This analysis almost completely parallels the analysis of the CS equation, but with one minor change: Instead of (\ref{gl8.2}) we now have:
\begin{equation}
\label{gl9.19}
  0 = \xi_A \partial_{\xi_A} {\cal S} (\Gamma ) =
  s_{\Gamma} (\xi_A \partial_{\xi_A} \Gamma ) - \chi_A \partial_{\xi_A} \Gamma
\end{equation}
However, differentiating the ST identity with respect to $\chi_A$ we find
\begin{equation}
\label{gl9.20}
  s_{\Gamma} (\partial_{\chi_A} \Gamma )
  = \partial_{\xi_A} \Gamma
\end{equation}
Hence the action principle, (\ref{gl9.19}) and (\ref{gl9.20}) imply that
\begin{equation}
\label{gl9.21}
  (\xi_A \partial_{\xi_A} + \chi_A \partial_{\chi_A} ) \Gamma  =
  \Delta_{\xi_A} \cdot \Gamma \; \; \; ,  
\end{equation}
where $\Delta_{\xi_A} \cdot \Gamma$ is a BRS symmetric insertion. But due to
\begin{equation}
\label{gl9.21a}
  s_{\Gamma} (\xi_A \partial_{\chi_A} \Gamma ) =
  (s_{\Gamma} \xi_A ) \partial_{\chi_A} \Gamma + \xi_A s_{\Gamma} (\partial_{\chi_A} \Gamma ) =
  \chi_A \partial_{\chi_A} \Gamma + \xi_A \partial_{\xi_A} \Gamma
\end{equation}
$\Delta_{\xi_A} \cdot \Gamma$ also has to be a BRS variation and hence only BRS variations can contribute to $\Delta_{\xi_A} \cdot \Gamma$.
From here on the discussion is completely analogous to the discussion of the CS equation; we skip the details and just present the result:
\begin{eqnarray}
\label{gl9.22}
  & \left( \xi_A \tilde{\partial}_{\xi_A} + \chi_A \partial_{\chi_A} +
  \tilde{\beta}_{\xi}^{\xi_A} \tilde{\partial}_{\xi}
  - \tilde{\gamma}_s^{\xi_A} {\cal N}_s^{(\chi , \chi_A )} -
  \tilde{\hat{\gamma}}_s^{\xi_A} \hat{\cal N}_s^{(\chi , \chi_A )} -
  \tilde{\bar{\gamma}}_s^{\xi_A} \bar{\cal N}_s^{(\chi , \chi_A )} -
  \tilde{\gamma}_B^{\xi_A} {\cal N}_B^{(\chi , \chi_A )}
   \right) \Gamma & \nonumber \\
  & = \; \; \; (\chi \partial_{\xi} + \chi_A \partial_{\xi_A} ) \; z^{-1} \; \xi_A \partial_{\xi_A} \int
  z Y_1 (\varphi_1 - \hat{\xi}_A \frac{m}{e} ) &
\end{eqnarray}
In (\ref{gl9.22}) $\xi_A \tilde{\partial}_{\xi_A}$ is the $W^{gen}$-symmetric extension of
$\xi_A \partial_{\xi_A}$ and given by: 
\begin{eqnarray}
\label{gl9.22a}
  \xi_A \tilde{\partial}_{\xi_A} & \equiv & \xi_A \partial_{\xi_A} -
  z^{-1} \; \xi_A \partial_{\xi_A} \int z \left\{
  (\varphi_1 - \hat{\xi}_A \frac{m}{e} ) \frac{\delta}{\delta \varphi_1} +
  (\hat{\varphi}_1 - \xi_A \frac{m}{e} ) \frac{\delta}{\delta \hat{\varphi}_1} -
  Y_1 \frac{\delta}{\delta Y_1} + q_1 \frac{\delta}{\delta q_1} \right\} \nonumber \\
  & & - (\chi \partial_{\xi} + \chi_A \partial_{\xi_A} ) \; z^{-1} \; \xi_A \partial_{\xi_A} \int z
  (\hat{\varphi}_1 - \xi_A \frac{m}{e} ) \frac{\delta}{\delta q_1}
\end{eqnarray}
Introducing the ``real'' $\beta$- and $\gamma$-functions like in (\ref{gl9.11a}) the test of (\ref{gl9.22}) on the gauge condition (\ref{gl4.2}) yields:
\begin{eqnarray}
\label{gl9.22b}
  \gamma_B^{\xi_A} & = & 0 \nonumber \\
  \beta_{\xi}^{\xi_A}  \; \; = \; \; 2 \gamma_B^{\xi_A} & = & 0 \\
  - \gamma_s^{\xi_A} - \hat{\gamma}_s^{\xi_A} - \gamma_B^{\xi_A} \; \; = \; \;
  (\xi_A \partial_{\xi_A} + \beta_{\xi}^{\xi_A} \xi \partial_{\xi} ) \mbox{ln} z &
  \Leftrightarrow & \hat{\gamma}_s^{\xi_A} = - \gamma_s^{\xi_A} - \xi_A \partial_{\xi_A}
  \mbox{ln} z \nonumber
\end{eqnarray}
Again, we can separate in (\ref{gl9.22}) the hard and soft breaking on the left and right hand side;
thereby using (\ref{gl9.22b}) we end up with the following form, which for brevity we only give for all external fields set equal to zero:
\begin{eqnarray}
\label{gl9.22c}
  & \left. \left( \xi_A \partial_{\xi_A} + \chi_A \partial_{\chi_A} - \gamma^{\xi_A}_s N_s -
     \hat{\gamma}^{\xi_A}_s \hat{N}_s - \bar{\gamma}^{\xi_A}_s \bar{N}_s -
     \xi_A \partial_{\xi_A} \mbox{ln} z \int
     \varphi_1 \frac{\delta}{\delta \varphi_1} \right) \Gamma \right|_{ext. f. \equiv 0} & \nonumber \\
  & = \; \;  - \frac{m}{e} \left. \int \left\{
            \alpha \frac{\delta}{\delta \varphi_1} +
	    \xi_A (1 - \gamma^{\xi_A}_s ) \frac{\delta}{\delta \hat{\varphi}_1} +
	    (\chi \partial_{\xi} + \chi_A \partial_{\xi_A} ) \xi_A (1 - \gamma^{\xi_A}_s ) 
	    \frac{\delta}{\delta q_1} \right\} \Gamma \right|_{ext. f. \equiv 0} &
\end{eqnarray}
with
\begin{equation}
\label{gl9.22d}
  \alpha = - \hat{\gamma}^{\xi_A}_s \hat{\xi}_A + \bar{\gamma}^{\xi_A}_s \xi_A +
                \xi_A \partial_{\xi_A} \hat{\xi}_A = x \xi_A + {\cal O} (\hbar )
\end{equation}		

\newsubsection{RG equation}

The derivation of the RG equation once more starts with the action principle
\begin{equation}
\label{gl9.51}
  \kappa \partial_{\kappa} \Gamma = \Delta_{\kappa} \cdot \Gamma
\end{equation}
which tells us that $\Delta_{\kappa} \cdot \Gamma$ has to be an insertion of dimension less than or equal to four, invariant under charge conjugation and in addition BRS symmetric due to
(\ref{gl8.2}).  In order to arrive at a more convenient form of the RG equation we now introduce
a new set of BRS symmetric operators (see also \cite{K2}) representing the two- and three-dimensional BRS symmetric classical field polynomials, i.e. instead of (\ref{gl8.19}) and $\int \delta_{\hat{\varphi}_0}$ we are going to use:
\begin{eqnarray}
\label{gl9.52}
  m \partial_m & , & f_A \partial_{\xi_A} \Gamma + [(\chi \partial_{\xi} + \chi_A \partial_{\xi_A} )
                               f_A ] \partial_{\chi_A} \Gamma = s_{\Gamma} (f_A \partial_{\chi_A}
			        \Gamma ) \; \; \; , \\
  \int \frac{\delta}{\delta \hat{\varphi}_0} & , &
  \tilde{f} \int \left\{ B \hat{\varphi}_2 - \bar{c} q_2 \right\} +
  (\chi \partial_{\xi} + \chi_A \partial_{\xi_A} ) \tilde{f} \int \bar{c} \hat{\varphi}_2 =
  s_{\Gamma} (\tilde{f} \int \bar{c} \hat{\varphi}_2 ) \nonumber
\end{eqnarray}
Hence according to  BRS invariance alone, $\Delta_{\kappa} \cdot \Gamma$ can be decomposed into a sum of the BRS symmetric operators (\ref{gl9.52}), (\ref{gl8.20}) and (the remaining four-dimensional operators in) (\ref{gl8.17}):
\begin{eqnarray}
\label{gl9.53}
  \kappa \partial_{\kappa} \Gamma & = & \!\! \! \! \left( - \beta^{\kappa}_m m \partial_m -
  \tilde{\beta}^{\kappa}_{\xi_A} (f_A \partial_{\xi_A} + [(\chi \partial_{\xi} + \chi_A \partial_{\xi_A} )
  f_A ] \partial_{\chi_A} +
  \alpha^{\kappa}_{inv} \int \delta_{\hat{\varphi}_0} -
  \beta^{\kappa}_{m_H} m_H \partial_{m_H} \right. \nonumber \\
  & & \! \! \! \! \left.  - \beta^{\kappa}_e e \partial_e
  + \gamma^{\kappa}_A N_A + \tilde{\gamma}^{\kappa}_B {\cal N}^{(\chi, \chi_A )}_B +
  \sum_{i=1}^2 \left\{
  \tilde{\gamma}^{\kappa}_{s,i} {\cal N}^{(\chi, \chi_A)}_{s,i} +
  \tilde{\bar{\gamma}}^{\kappa}_{s,i} \bar{\cal N}^{(\chi, \chi_A)}_{s,i} +
  \tilde{\hat{\gamma}}^{\kappa}_{s,i} \hat{\cal N}^{(\chi, \chi_A)}_{s,i} \right \} \right. \nonumber \\
  & & \! \! \! \! \left. - \tilde{\beta}^{\kappa}_{\xi} (f_{\xi} \partial_{\xi} +
  [(\chi \partial_{\xi} + \chi_A \partial_{\xi_A} ) f_{\xi} ] \partial_{\chi} \right) \Gamma +
  \tilde{\gamma}^{\kappa} s_{\Gamma} (\tilde{f}_3 \int \bar{c} \hat{\varphi}_2 ) +
  \tilde{\tilde{\gamma}}^{\kappa} s_{\Gamma} 
  (\tilde{f}_4 \int \bar{c} \hat{\varphi}_1 \hat{\varphi}_2 )
\end{eqnarray}
Differentiating (\ref{gl9.53}) with respect to $\varphi_1$, setting all fields equal to zero and making use of the normalization condition $\Gamma_{\varphi_1} = 0 $ (\ref{vacnorm}) it immeadiately follows that:
\begin{equation}
\label{gl9.54}
  \alpha^{\kappa}_{inv} \equiv 0
\end{equation}
With this result in mind three further tests of (\ref{gl9.53}) on the physical normalization conditions (\ref{massnorm}) concerning the mass normalizations of the Higgs, the vector and the ghost
directly imply
\begin{equation}
\label{gl9.55}
  \beta^{\kappa}_{m_H} \equiv 0 \; \; \; , \; \; \;
  \beta^{\kappa}_m \equiv 0 \; \; \; , \; \; \;
  \tilde{\beta}^{\kappa}_{\xi_A} \equiv 0
\end{equation}
to all orders of perturbation theory. Therefore due to the physical normalization conditions the first line of the r.h.s. of (\ref{gl9.53}) is absent and no $\beta$-function in connection with a (physical) mass appears in the RG equation.\\
In order to conclude the derivation of the RG equation we now have to exploit rigid invariance of the
theory: To this end we first apply $W^{gen}$ (\ref{gl6.7}) to the RG equation (\ref{gl9.53}) and then also use the rigid WI (\ref{gl6.10}):
\begin{equation}
\label{gl9.56}
  W^{gen} \kappa \partial_{\kappa} \Gamma =
  [W^{gen} , \kappa \partial_{\kappa} ] \Gamma + \kappa \partial_{\kappa} W^{gen} \Gamma =
  - (\kappa \partial_{\kappa} W^{gen} ) \Gamma + \kappa \partial_{\kappa} 
  (\chi \Delta_{br} + \chi_A \Delta_{br_A} )
\end{equation}
This leads after some calculation to the final form of the RG equation:
\begin{eqnarray}
\label{gl9.57}
  & & \left( \kappa \partial_{\kappa} + \beta^{\kappa}_e e \partial_e +
         \beta^{\kappa}_{\xi} \xi \partial_{\xi} -
	 \gamma^{\kappa}_A N_A - \gamma^{\kappa}_B N_B -
	 \gamma^{\kappa}_s N_s - \hat{\gamma}^{\kappa}_s \hat{N}_s -
	 \bar{\gamma}^{\kappa}_s \bar{N}_s \right. \nonumber \\
  & & \left. - \gamma^{\kappa}_1 \int \left\{
         \varphi_1 \frac{\delta}{\delta \varphi_1} - Y_1 \frac{\delta}{\delta Y_1} +
	 \hat{\varphi}_1 \frac{\delta}{\delta \hat{\varphi}_1} + q_1 \frac{\delta}{\delta q_1} \right\}
	 \right. \nonumber \\
  & & \left. - (\chi \partial_{\xi} + \chi_A \partial_{\xi_A} ) \left[ \int \left\{
         (\hat{\gamma}^{\kappa}_s + \gamma^{\kappa}_1 )
	 \hat{\varphi}_1 \frac{\delta}{\delta q_1} +
	 \hat{\gamma}^{\kappa}_s \hat{\varphi}_2 \frac{\delta}{\delta q_2} +
	 \gamma^{\kappa}_B \bar{c} \frac{\delta}{\delta B} \right\} -
	 \beta^{\kappa}_{\xi} \xi \partial_{\chi} \right] \right) \Gamma \nonumber \\
  & = & \bar{\gamma}^{\kappa}_s \int \left\{ q_1 Y_1 + q_2 Y_2 \right\} \\
  & & + (\chi \partial_{\xi} + \chi_A \partial_{\xi_A} ) \int \left\{
         (\gamma^{\kappa}_s + \gamma^{\kappa}_1 ) Y_1 \varphi_1 +
	 \gamma^{\kappa}_s Y_2 \varphi_2 +
	 \bar{\gamma}^{\kappa}_s (Y_1 \hat{\varphi}_1 + Y_2 \hat{\varphi}_2 ) \right\} \nonumber
\end{eqnarray}
with
\begin{equation}
\label{gl9.58}
  \gamma^{\kappa}_1 = (\kappa \partial_{\kappa} +
  \beta^{\kappa}_e e \partial_e +
  \beta^{\kappa}_{\xi} \xi \partial_{\xi} ) \mbox{ln} z
\end{equation}
In (\ref{gl9.57}) we have already introduced the full $\beta$- and $\gamma$-functions of the RG equation like in (\ref{gl9.11a}). Again, our analysis shows that the $\beta$-function $\beta^{\kappa}_e$ and the anomalous dimension $\gamma^{\kappa}_A$ have to be $\xi$- and
$\xi_A$-independent to all orders of the loop expansion.\\
Additionally, rigid invariance (\ref{gl9.56}) also imposes two restrictions for the coefficient functions
of the RG equation:
\begin{eqnarray}
\label{gl9.59}
  (\kappa \partial_{\kappa} + \beta^{\kappa}_{\xi} \xi \partial_{\xi} +
   \beta^{\kappa}_e e \partial_e ) (z \hat{\xi}_A \frac{m}{e} ) & = &
  - z \xi_A \frac{m}{e} \bar{\gamma}^{\kappa}_s 
  - z \hat{\xi}_A \frac{m}{e} \gamma^{\kappa}_s \nonumber \\
  \beta^{\kappa}_e - \hat{\gamma}^{\kappa}_s & = &
  (\kappa \partial_{\kappa} + \beta^{\kappa}_{\xi} \xi \partial_{\xi} +
   \beta^{\kappa}_e e \partial_e ) \mbox{ln} z
\end{eqnarray}
Some further information about the coefficient functions results from testing the RG equation on the
gauge condition (\ref{gl4.2})\footnote{When deriving (\ref{gl9.60}) we make use of (\ref{gl9.59}).}:
\begin{eqnarray}
\label{gl9.60}
  & \gamma^{\kappa}_B = - \gamma^{\kappa}_A & \\
  & \beta^{\kappa}_{\xi} = 2 \gamma^{\kappa}_B = - 2 \gamma^{\kappa}_A & \nonumber \\
  & \gamma^{\kappa}_s = - \gamma^{\kappa}_B = \gamma^{\kappa}_A \nonumber
\end{eqnarray}
Hence also $\gamma^{\kappa}_B, \gamma^{\kappa}_s$ and $\beta^{\kappa}_{\xi}$ are fully gauge
parameter independent.\\
Finally, one further relation emerges from the validity of the local WI (\ref{gl7.14}) and the normalization condition for the coupling (\ref{gl7.16}):
\begin{equation}
\label{gl9.61}
  \gamma^{\kappa}_A = \beta^{\kappa}_e
\end{equation}
Therefore, there is only one independent coefficient function appearing in the RG equation, namely
the $\beta$-function $\beta^{\kappa}_e$, which has to be determined by an explicit calculation.

\newsection{Conclusions}

In the present paper we have examined the renormalization of the Abelian Higgs model including
BRS variations of all the gauge parameters. The advantage of such an extended procedure (when compared to the usual one) is due to the fact that this procedure also yields full information about the gauge parameter dependence of 1-PI Green functions automatically and in an easily manageable way and therefore prohibits (just by construction) a wrong adjustment of counterterms which in turn would spoil the gauge parameter independence of the S-matrix. In the usual construction (i.e. without introducing BRS transforming gauge parameters) such a simple guiding principle is missing and it is a quite troublesome and heavily controllable task to adjust the counterterms correctly.\\
In this context we have shown that the normalization conditions needed in order to fix the free parameters of the theory cannot be chosen arbitrarily but instead have to respect the restrictions dictated by the enlarged ST identity. Especially we have proven that the physical on-shell normalization conditions are in complete agreement with those restrictions. Furthermore, the method of BRS varying gauge parameters yields a well handleable tool for controlling the range of ``good'' normalization conditions, i.e. normalization conditions, which are not in contradiction with the enlarged ST identity.\\
Some further results of the algebraic method we find interesting, too:\\
The enlarged ST identity also allowed us to show that the transversal part of the vector 2-point function has to be completely gauge parameter-independent to all orders of perturbation theory.\\
In the course of proving the local WI we found the $\xi$- and $\xi_A$-independence of the overall normalization factor of the matter transformations, a result, which gave rise to an alternative and elegant possibility for fixing the coupling, namely by requiring the local WI to be exact to all orders.\\
Finally, we derived the Callan-Symanzik and the renormalization group equation of the Abelian Higgs model thereby showing among other things that the $\beta$-functions $\beta_e^{(\kappa )},
\beta_{m_H}$ and $\beta_{\xi}^{(\kappa)}$ as well as the anomalous dimensions $\gamma_A^{(\kappa )}, \gamma_B^{(\kappa)}$ and $\gamma_s^{\kappa}$ have to be fully gauge parameter-independent to all orders of the perturbative expansion.\\
The examination of the Abelian Higgs model, chosen as the simplest example of a gauge theory with spontaneous breakdown of symmetry, thus clearly shows of what kind the considerations have to be and yields a hint what kind of results could possibly be expecxted when the general algebraic method will be applied to more complicated, physical, models, especially to the standard model of electroweak interactions.\\[1cm]
{\it Acknowledgements} \hspace{0.4cm} The authors would like to thank K. Sibold and E. Kraus for
numerous helpful discussions and a critical reading of the manuscript.\\[2cm]
\renewcommand{\theequation}{A.\arabic{equation}}
\setcounter{equation}{0}
\noindent
{\large \bf Appendix A}\\[0.5cm]
In the course of looking for the most general classical solution of the 
$\chi$- and $\chi_A$-enlarged ST identity (\ref{gl4.1}) the most general solution of
the gauge condition (\ref{gl4.2}) and the ordinary (that is $\chi$- and $\chi_A$-independent)
ST identity
\begin{equation}
\label{gla.1}
{\cal S} (\hat{\Gamma}) = \int \left\{
  \partial_{\mu} c \; \frac{\delta \hat{\Gamma}}{\delta A_{\mu}}
\; + \; B \; \frac{\delta \hat{\Gamma}}{\delta \bar{c}}
\; + \; \frac{\delta \hat{\Gamma}}{\delta \underline{Y}} \;
        \frac{\delta \hat{\Gamma}}{\delta \underline{\varphi}}
\; + \; \underline{q} \;
        \frac{\delta \hat{\Gamma}}{\delta \underline{\hat{\varphi}}}
\right\} = 0
\end{equation}
is needed. This solution was constructed in \cite{KS}, and we just present the result here:
\begin{equation}
\label{gla.2}
\hat{\Gamma}_{cl}^{gen} = \Lambda (A_{\mu}, \bar{\varphi}_1, \bar{\varphi}_2 )
\; + \; \Gamma_{g.f.} \; + \; \Gamma_{\phi \pi} \; + \; \Gamma_{e.f.} \; ,
\end{equation}
with
\begin{equation}
\label{gla.3}
\bar{\varphi}_i = \varphi_i - x_i \hat{\varphi}_i \; , \; i = 1,2 \; .
\end{equation}
The part
 $\Lambda = \Lambda (A_{\mu}, \bar{\varphi}_1, \bar{\varphi}_2 )$ 
describing the gauge field $A_{\mu}$ and matter fields $\varphi_i$ is given by:
\begin{eqnarray}
\label{gla.4}
\Lambda & = &
  \int \left\{ -\; \frac{z_A}{4} F_{\mu \nu} F^{\mu \nu} \; + \;
  \frac{1}{2} z_1 (\partial_{\mu} \bar{\varphi}_1 )
                  (\partial^{\mu} \bar{\varphi}_1 ) \; + \;
\frac{1}{2} z_2 (\partial_{\mu} \bar{\varphi}_2 )
                (\partial^{\mu} \bar{\varphi}_2 ) \right. \nonumber \\
& & \left. + \; z_e e \sqrt {z_1} \sqrt{z_2} \sqrt{z_A} \left( 
                (\partial_{\mu} \bar{\varphi}_1 ) \bar{\varphi}_2 \; - \;
                \bar{\varphi}_1 (\partial_{\mu} \bar{\varphi}_2 ) \right)
                A^{\mu} \; + \;
\frac{1}{2} z_e^2 {e}^2 z_A  (z_1 \bar{\varphi}_1^2 \; + \;
                               z_2 \bar{\varphi}_2^2) A_{\mu} A^{\mu}
\right. \nonumber \\
& & \left. + \; \frac{1}{2} z_m m^2 z_A  A_{\mu} A^{\mu} \; - \;
    \sqrt{ z_2}\sqrt{z_m} m \sqrt{z_A}
(\partial_{\mu} \bar{\varphi}_2 ) A^{\mu} \; + \;
z_e e \sqrt{ z_m} m \sqrt {z_1} z_A
\bar{\varphi}_1 A_{\mu} A^{\mu} \right. \nonumber \\
& & \left. + \; \frac{1}{2} \mu^2
    (z_1 \bar{\varphi}_1^2 + 2 \sqrt{z_1} 
\frac {\sqrt{z_m} m}{{z_e} e} \bar{\varphi}_1 +
     z_2 \bar{\varphi}_2^2) \right. \nonumber \\
& & \left. \; - 
\frac 18 \frac{z_{m_H}m_H^2}{z_m m^2}z_e^2 e^2 (z_1 \bar{\varphi}_1^2 + 
2 \sqrt{z_1} \frac {\sqrt{z_m} m}{{z_e} e} 
 \bar{\varphi}_1 +
z_2 \bar{\varphi}_2^2)^2 \right\}
\end{eqnarray}
The gauge fixing part $\Gamma_{g.f.}$ immediately results from integrating
the gauge condition (\ref{gl4.2}):
\begin{equation}
\label{gla.5}
\Gamma_{g.f.} = \int \left\{ \frac{1}{2} \xi B^2 + 
  B \partial A - 
e B \left[ (\hat{\varphi}_1 - \xi_A \frac{m}{e} ) \varphi_2 -
            \hat{\varphi}_2 (\varphi_1 - \hat{\xi}_A \frac{m}{e}) \right]
\right\}
\end{equation}
For the remaining two parts, the external field part $\Gamma_{e.f.}$ and the
$\phi \pi$-part $\Gamma_{\phi \pi}$, one gets
\begin{equation}
\label{gla.6}
\Gamma_{e.f.} = \int \{ Y_1 (-{e}z_e \sqrt {\frac{z_2}{z_1}} 
\sqrt {z_A} \bar{\varphi}_2 c +
                                  x_1 q_1) \; + \;
Y_2 ({e}z_e \sqrt {\frac{z_1}{z_2}} \sqrt{z_A} (\bar{\varphi}_1 + 
\frac {\sqrt{z_m} m}{\sqrt{z_1} {z_e} e} ) c + x_2 q_2) \}
\end{equation}
and
\begin{eqnarray}
\label{gla.7}
\Gamma_{\phi \pi} & = & \int \{
  - \bar{c} \Box c \; + \; e\bar{c} (q_1 \varphi_2 - q_2 (\varphi_1 -
                               \hat{\xi}_A \frac{m}{e})) \nonumber \\
& & + \; e \bar{c} (\hat{\varphi}_1 - \xi_A \frac{m}{e})
                   (z_e e \sqrt{\frac{z_1}{z_2}} \sqrt{z_A} 
 (\bar{\varphi}_1 + \frac{\sqrt{z_m} m}{\sqrt{z_1} z_e e} ) c + x_2 q_2 )
    \nonumber \\
& & - \; e \bar{c} \hat{\varphi}_2 
 (-z_e e \sqrt{\frac{z_2}{z_1}} \sqrt{z_A} \bar{\varphi}_2 c
                                           + x_1 q_1) \} \; .
\end{eqnarray}
The free parameters in the general solution of the ST identity (\ref{gla.1}) 
are the wave function normalizations $z_1,
z_2$ and $z_A$, the mass renormalizations of the vector and the Higgs-particle, i.e.
$z_m, z_{m_H}$, the coupling renormalization $z_e$, the parameters $x_1, x_2$, the parameter $\mu$, the gauge parameters $\xi, \xi_A$ and the parameter $\hat{\xi}_A$. These parameters
are not prescribed by the ST identity (\ref{gla.1}) and therefore have
to be fixed by appropriate normalization conditions to all orders (see section 2).\\[1cm]
\renewcommand{\theequation}{B.\arabic{equation}}
\setcounter{equation}{0}
\noindent
{\large \bf Appendix B}\\[0.5cm]
The solution of (\ref{gl8.15}) was given in \cite{KS}; first we present a list of all
terms of dimension less than or equal to {\it three} which contribute to $\hat{\Delta}_{\lambda}^0$:
\begin{eqnarray}
\label{glB.1}
  \displaystyle\int \displaystyle\frac{\hat{\Gamma}_{cl}}{\delta \varphi_1} & = &
     s_{\hat{\Gamma}_{cl}}^{\chi = 0 = \chi_A} \int Y_1 \; \; \; , \nonumber \\
  \displaystyle\int \displaystyle\frac{\delta \hat{\Gamma}_{cl}}{\delta \hat{\varphi}_1} & = &
     s_{\hat{\Gamma}_{cl}}^{\chi = 0 = \chi_A} \int \left\{
     -x Y_1 - e \bar{c} \bar{\varphi}_2 \right\} \; \; \; , \nonumber \\
  \int \left\{ B \hat{\varphi}_2 - \bar{c} q_2 \right\} & = &
      s_{\hat{\Gamma}_{cl}}^{\chi = 0 = \chi_A} \int \bar{c} \hat{\varphi}_2 
\end{eqnarray}
and
\begin{displaymath}
  \int \left\{ z_1 \bar{\varphi}_1^2 + 2 z_1 v \bar{\varphi}_1 + z_2 \bar{\varphi}_2^2 \right\}
\end{displaymath}
In order to have a proper definition of this last invariant in higher orders we are forced to
introduce a further external field $\hat{\varphi}_0$ of dimension two, even under charge
conjugation and invariant under BRS and rigid transformations, which couples to this invariant.
Therefore the above BRS symmetric term is replaced by:
\begin{equation}
\label{glB.2}
  \int \frac{\delta \hat{\Gamma}_{cl}}{\delta \hat{\varphi}_0}
\end{equation}
The {\it four}-dimensional BRS symmetric terms contributing to $\hat{\Delta}_{\lambda}^0$ are
given by ($i = 1,2$):
\begin{eqnarray}
\label{glB.3}
  \displaystyle\int \left\{ \varphi_i \displaystyle\frac{\delta}{\delta \varphi_i} - 
     Y_i \displaystyle\frac{\delta}{\delta Y_i} \right\}
     \hat{\Gamma}_{cl} & = &
     s_{\hat{\Gamma}_{cl}}^{\chi = 0 = \chi_A} \int Y_i \varphi_i \; \; \; , \nonumber \\
  \displaystyle\int \left\{ \hat{\varphi}_i 
      \displaystyle\frac{\delta \hat{\Gamma}_{cl}}{\delta \varphi_i} - Y_i q_i \right\} & = &
     s_{\hat{\Gamma}_{cl}}^{\chi = 0 = \chi_A} \int Y_i \hat{\varphi}_i \; \; \; , \nonumber \\
  \displaystyle\int \left\{ B \displaystyle\frac{\delta}{\delta B} + 
      \bar{c} \displaystyle\frac{\delta}{\delta \bar{c}} \right\}
      \hat{\Gamma}_{cl} & = &
     s_{\hat{\Gamma}_{cl}}^{\chi = 0 = \chi_A} \displaystyle\int \bar{c} 
     \displaystyle\frac{\delta \hat{\Gamma}_{cl}}{\delta B}
     \; \; \; , \nonumber \\
  \displaystyle\int \left\{ \hat{\varphi}_i \displaystyle\frac{\delta}{\delta \hat{\varphi}_i} + 
     q_i \displaystyle\frac{\delta}{\delta q_i} \right\}
     \hat{\Gamma}_{cl} & = &
     s_{\hat{\Gamma}_{cl}}^{\chi = 0 = \chi_A} \displaystyle\int \hat{\varphi}_i
     \displaystyle\frac{\delta \hat{\Gamma}_{cl}}{\delta q_i} \; \; \; , \nonumber \\
  \int \left\{ B \hat{\varphi}_1 \hat{\varphi}_2 - \bar{c} q_1 \hat{\varphi}_2 -
     \bar{c} \hat{\varphi}_1 q_2 \right\} & = &
     s_{\hat{\Gamma}_{cl}}^{\chi = 0 = \chi_A} \int \bar{c} \hat{\varphi}_1 \hat{\varphi}_2 \; \; \; ,
     \nonumber \\
  \xi \partial_{\xi} \hat{\Gamma}_{cl} & = &
     s_{\hat{\Gamma}_{cl}}^{\chi = 0 = \chi_A} \xi Q \; \; \; , 
\end{eqnarray}
and
\begin{equation}
\label{glB.4}
  \int \left\{ A \frac{\delta}{\delta A} + c\frac{\delta}{\delta c} \right\} \hat{\Gamma}_{cl} \; \; \; ,
  \; \; \; m_H \partial_{m_H} \hat{\Gamma}_{cl} \; \; \; , \; \; \;
  e \partial_e \hat{\Gamma}_{cl}
\end{equation}
Please note that due to (\ref{gl8.14}) the coefficients with which the terms in (\ref{glB.2}) and
(\ref{glB.4}) appear in $ \hat{\Delta}_{\lambda}^0$ are independent of both $\xi$ and $\xi_A$.\\[1cm]
\renewcommand{\theequation}{C.\arabic{equation}}
\setcounter{equation}{0}
\noindent
{\large \bf Appendix C}\\[0.5cm]
In this appendix we present the $W^{gen}$-symmetric extensions of the BRS invariant insertions
\begin{equation}
\label{glC.1}
  m_H \partial_{m_H} \; \; , \; \;
  e \partial_e \; \; , \; \;
  f_{\xi} \partial_{\xi} + (\chi \partial_{\xi} + \chi_A \partial_{\xi_A} ) f_{\xi} \partial_{\chi} \; \; .
\end{equation}
Just in order to compactify the notation in the formulae below we introduce two $\xi$- and $\xi_A$-independent factors $f_H$ and $f_e$ multiplying $m_H \partial_{m_H}$ and $e \partial_e$, respectively. (These factors have to be independent of $\xi$ and $\xi_A$ due to the results of section 8.) Next we define ($i = H, e, \xi$):
\begin{eqnarray}
\label{glC.2}
  \nabla_i & = & m_H \partial_{m_H} , e \partial_e , \partial_{\xi} \; \; \; , \\
\label{glC.3}
  \hat{\nabla}_i & = &  - f_i \frac{1}{z} \nabla_i  \int z \left\{
  (\varphi_1 - \hat{\xi}_A \frac{m}{e} ) \frac{\delta}{\delta \varphi_1} - Y_1 \frac{\delta}{\delta Y_1} +
  (\hat{\varphi}_1 - \xi_A \frac{m}{e} ) \frac{\delta}{\delta \hat{\varphi}_1} +
  q_1 \frac{\delta}{\delta q_1} \right\} \nonumber \\
  & & - (\chi \partial_{\xi} + \chi_A \partial_{\xi_A} ) f_i \frac{1}{z} \nabla_i \int z
  (\hat{\varphi}_1 - \xi_A \frac{m}{e} ) \frac{\delta}{\delta q_1} \; \; \; , \\
\label{glC.4}
  \hat{\hat{\nabla}}_i \Gamma & = & - (\chi \partial_{\xi} + \chi_A \partial_{\xi_A} ) f_i \frac{1}{z}
  \nabla_i \int z Y_1 (\varphi_1 - \hat{\xi}_A \frac{m}{e} )
\end{eqnarray}
The $W^{gen}$-symmetric extensions of the operators in (\ref{glC.1}) then are given by:
\begin{eqnarray}
\label{glC.5}
  f_H m_H \partial_{m_H} \Gamma & \rightarrow & f_H m_H \tilde{\partial}_{m_H} \Gamma \equiv
  f_H m_H \partial_{m_H} \Gamma + \hat{\nabla}_H \Gamma + \hat{\hat{\nabla}}_H \Gamma
  \; \; \; , \nonumber \\
  f_e e \partial_e \Gamma & \rightarrow & f_e e \tilde{\partial}_e \Gamma \equiv
  f_e e \partial_e \Gamma + \hat{\nabla}_e \Gamma + \hat{\hat{\nabla}}_e \Gamma
  \; \; \; , \\
  f_{\xi} \partial_{\xi} \Gamma + [(\chi \partial_{\xi} + \chi_A \partial_{\xi_A} ) 
  f_{\xi} ] \partial_{\chi} \Gamma
  & \rightarrow & \tilde{\partial}_{\xi} \Gamma \equiv
  f_{\xi} \partial_{\xi} \Gamma + [(\chi \partial_{\xi} + \chi_A \partial_{\xi_A} ) 
  f_{\xi} ] \partial_{\chi} \Gamma
  + \hat{\nabla}_{\xi} \Gamma + \hat{\hat{\nabla}}_{\xi} \Gamma \nonumber
\end{eqnarray}
\vspace{2cm}
\newpage
{

}
\end{document}